\documentclass[12pt,preprint]{aastex}

\newcommand{\ldot}{L$_{\odot}$}
\newcommand{\um}{$\mu$m~}
\newcommand{\ums}{$\mu$m}

\def\kmsMpc{\ifmmode {\rm\,km\,s^{-1}\,Mpc^{-1}}\else
    ${\rm\,km\,s^{-1}\,Mpc^{-1}}$\fi}

\shorttitle{Average Spitzer Spectra for Galaxies}
\shortauthors{Weedman and Houck}

\begin{document}
\title{Average Infrared Galaxy Spectra From Spitzer Flux Limited Samples} 

\author{Daniel W. Weedman\altaffilmark{1} and James R. Houck\altaffilmark{1}}
 
\altaffiltext{1}{Astronomy Department, Cornell University, Ithaca, NY 14853; dweedman@isc.astro.cornell.edu}

\begin{abstract}
  
The mid-infrared spectroscopic analysis of a flux-limited sample of galaxies with f$_{\nu}$(24\ums) $>$ 10\,mJy is presented.  Sources observed are taken from the $Spitzer$ First Look Survey (FLS) catalog and from the NOAO Deep Wide-Field Survey region in Bootes (NDWFS).  The spectroscopic sample includes 60 of the 100 sources in these combined catalogs having f$_{\nu}$(24\ums) $>$ 10\,mJy.  New spectra from the $Spitzer$ Infrared Spectrograph are presented for 25 FLS sources and for 11 Bootes AGN; these are combined with 24 Bootes starburst galaxies previously published to determine the distribution of mid-infrared spectral characteristics for the total 10\,mJy sample.  Sources have 0.01 $<$ z $<$ 2.4 and 41.8 $<$ log $\nu$L$_{\nu}$ (15$\mu$m) $<$ 46.2 (ergs s$^{-1}$).  Average spectra are determined as a function of luminosity; lower luminosity sources (log $\nu$L$_{\nu}$(15$\mu$m) $<$ 44.0) are dominated by PAH features and higher luminosity sources (log $\nu$L$_{\nu}$(15$\mu$m) $>$ 44.0 ) are dominated by silicate absorption or emission. We find that a rest frame equivalent width of 0.4\,\um for the 6.2\,\um PAH emission feature provides a well defined division between lower luminosity, "pure" starbursts and higher luminosity AGN or composite sources. Using the average spectra, fluxes f$_{\nu}$(24\ums) which would be observed with the $Spitzer$ MIPS are predicted as a function of redshift for sources with luminosities that correspond to the average spectra. AGN identical to those in this 10 mJy sample could be seen to z = 3 with f$_{\nu}$(24\ums) $>$ 1 mJy, but starbursts fall to f$_{\nu}$(24\ums) $<$ 1 mJy by z $\sim$ 0.5.  This indicates that substantial luminosity evolution of starbursts is required to explain the numerous starbursts found in other IRS results having f$_{\nu}$(24\ums) $\sim$ 1 mJy and z $\sim$ 2. 
\end{abstract}



\section{Introduction}

A full understanding of the nature and evolutionary characteristics of the extragalactic sources discovered at mid-infrared wavelengths by the Spitzer Space Telescope ($Spitzer$) requires comparison of the spectroscopic and photometric characteristics of these sources.  The astrophysical nature of individual sources can be understood from spectra obtained with The Infrared Spectrograph on $Spitzer$ (IRS, Houck et al. 2004).  If characteristic mid-infrared spectra for a variety of the sources are known, flux distributions of sources within photometric surveys with the Multiband Imaging Photometer for $Spitzer$ (MIPS, Rieke et al. 2004) can be compared with source count models \citep[e.g. ][]{pap04} to determine the evolution in the universe of dusty sources, primarily starbursts and active galactic nuclei (AGN).    

The fundamental method to determine representative mid-infrared spectral characteristics is to obtain spectra with the IRS for unbiased samples.  The primary limitation is that IRS spectra can only be obtained for significant numbers of sources with f$_{\nu}$(24\ums) $\ga$ 1 mJy.  Despite this limitation, it is crucial to determine as thoroughly as possible the spectral characteristics of well-defined, flux-limited samples.  In the present paper, we take an initial step toward this determination using a sample of sources discovered in MIPS surveys and defined only by f$_{\nu}$(24\ums) $>$ 10 mJy.

Our objective in observing this 10 mJy sample is to determine the range of spectral characteristics among sources defined only by discovery with $Spitzer$ based on mid-infrared flux.  The result allows a determination of how representative are spectra of the many sources observed with the IRS but chosen using various other selection criteria, such as optical spectral classification or prior knowledge of the source luminosities (e.g. the many available IRS spectra of Ultraluminous Infrared Galaxies). 

Previously, we published a catalog of 50 sources \citep{hou07} from the sample determined from the MIPS survey of 8.2 deg$^{2}$ within the Bootes field of the NOAO Deep Wide-Field Survey (NDWFS, Jannuzi et al. 1999), although only the spectra of the starburst sources were presented.  In the present paper, we present the complete spectral sample of the 25 extragalactic sources having f$_{\nu}$(24\ums) $>$ 10 mJy in the $Spitzer$ First Look Survey (FLS) point source catalog \citep{fad06}; we also present and discuss the spectra of the AGN within the Bootes sample. 

Within the combined Bootes and FLS surveys, there are a total of 100 extragalactic sources with f$_{\nu}$(24\ums) $>$ 10 mJy.  By combining new observations in our programs with existing observations in the $Spitzer$ archive from other programs, we have complete low-resolution IRS spectra for 35 of 50 Bootes sources having f$_{\nu}$(24\ums) $>$ 10 mJy and for all 25 FLS point sources having f$_{\nu}$(24\ums) $>$ 10 mJy.  The present paper is an analysis of the resulting 60 spectra of sources having f$_{\nu}$(24\ums) $>$ 10 mJy.  

Sources included within our sample were selected using only the flux limit criterion of f$_{\nu}$(24\ums) $>$ 10 mJy with no other selection criteria.  Our sample is not, however, a complete sample of all sources in Bootes and the FLS having f$_{\nu}$(24\ums) $>$ 10 mJy.  This is primarily because nearby, extended sources (z $\la$ 0.05) having previously known optical redshifts and optical spectral classifications (usually consistent with starbursts) were not observed with the IRS.  In particular, no IRS spectra were obtained for the 25 extragalactic sources having f$_{\nu}$(24\ums) $>$ 10 mJy in the FLS extended source catalog.  Six of these 25 sources are within NGC or IC galaxies; 24 of the 25 sources have optical redshifts in the National Extragalactic Database, with an average redshift of 0.0362. 

Because of the difference between a flux limited sample and a complete sample, we do not use the present flux limited sample to determine luminosity functions or to determine the quantitative fractions of sources which fall within various spectroscopic categories, such as starbursts or AGN.  Our primary objective is to obtain a census of sources covering a wide range of luminosities in order to determine mid-infrared spectral characterisitics as a function of luminosity.  For this objective, the most important goal is to include comparable numbers of sources within different luminosity bins.  It was not efficient, therefore, to obtain numerous IRS spectra of nearby starbursts having similar luminosities because the similarity among IRS spectra of such starbursts is well established \citep{bra06}.

\section{Observations}
\subsection{FLS Sample and New IRS Observations}

In the point source catalog of the FLS \citep{fad06}, there are 38 sources with f$_{\nu}$(24\ums) $>$ 10 mJy.  Of these, 13 are bright galactic stars.  The remaining 25 sources have been observed with the IRS: 19 are in our program 40038, 4 in archival program 20128 (G. Lagache, P.I.) one in archival program 20083 (M. Lacy, P.I.) and one in archival program 40539 (G. Helou, P.I.).  Characteristics and observational details of these 25 sources are given in Table 1. 

$Spitzer$ spectroscopic observations were made with the IRS\footnote{The IRS was a collaborative venture between Cornell
University and Ball Aerospace Corporation funded by NASA through the
Jet Propulsion Laboratory and the Ames Research Center.} Short Low module in
orders 1 and 2 (SL1 and SL2) and with the Long Low module in orders 1 and 2 (LL1 and
LL2), described in \citet{hou04}.  These give low resolution spectral
coverage from $\sim$5\,\um to $\sim$35\,\ums.  

For our new observations, sources were placed on
the slit by using the IRS peakup mode with the blue camera (13.5\um $<$ $\lambda$ $<$ 18.7\ums).  All images when the source was in one of the two nod positions on each
slit were coadded to obtain the image of the source spectrum.  The background which was subtracted was determined from coadded background images that added both nod positions having the source in the other slit (i.e., both nods on the LL1 slit when the source is in the LL2 slit produce LL1 images of background only).  
The difference between coadded source images minus coadded background images was used for the spectral extraction, giving two independent extractions of the spectrum for each order.  These independent spectra were compared to reject any highly outlying pixels in either
spectrum, and a final mean spectrum was produced.  

The extraction of one dimensional spectra from the two dimensional images was done with the SMART analysis package \citep{hig04}, beginning with the Basic Calibrated Data products, version 15, of the $Spitzer$ flux calibration pipeline.  Final spectra were boxcar-smoothed to the approximate resolution of the different IRS modules (0.2\,\um for SL1 and SL2, 0.3\,\um for LL2, and 0.4\,\um for LL1).  

Spectra for all FLS sources in Table 1 are illustrated among the spectra in Figures 1-4.  Spectra for FLS starbursts are in Figure 1 and are truncated at 20\,\um in the rest frame because of the absence of any significant features in the low resolution spectra beyond that wavelength; most Bootes starburst spectra were illustrated in \citet{hou07}.  Spectra for Bootes and FLS AGN in Table 3 are illustrated in Figures 2-4.  

Measured spectral parameters for all starbursts with IRS spectra from FLS and Bootes are summarized in Tables 2 and 3, and spectral parameters for all AGN are in Tables 4 and 5.  For purposes of classifying sources as starburst or AGN for inclusion in the Tables and Figures, sources with strong polycyclic aromatic hydrocarbon (PAH) features are defined as starbursts, and sources with silicate absorption or emission features are defined as AGN.  While such classifications could be further quantified, and while there are certainly composite sources, the systematic differences between spectra defined by this simple classification are obvious when comparing the starbursts in Figure 1 with the AGN in Figures 2-4.

\subsection{Sources with PAH features}

All of the sources from the Bootes and FLS samples which are characterised primarily by PAH emission are listed in Tables 2 and 3.  These Tables give measured fluxes and equivalent widths (EW) for PAH emission features and for the [NeIII] 15.56\,\um emission line;  continuum flux densities at 5.5\,\um and 15\,\um are also given.  Equivalent widths and fluxes are given for the 6.2$\mu$m and 11.3$\mu$m features; these are measured within SMART as single Gaussians on top of a linear continuum, using the continuum baseline between 5.5\,\um to 6.9\,\um for the 6.2\,\um feature, between 10.4\,\um to 12.2\,\um for the 11.3\,\um feature, and between 14.9\,\um to 16.2\,\um for the 15.6\,\um feature. As discussed by \citet{hou07} and by \citet{wee08}, the stronger 7.7$\mu$m feature is measured only by the flux density at the peak of the feature, which is a combination of feature strength and continuum strength, because of the large uncertainty in separating the underlying continuum from the PAH feature.  Most values for the Bootes sources in Tables 2 and 3 are reproduced from \citet{hou07}.  Results for all FLS sources and for archival Bootes sources in program 20113 (H. Dole, P.I.) derive from our new spectral extractions.  

The continuum strength which we choose for comparing source luminosities is measured by the flux density at rest frame 15\ums.  This wavelength is chosen because it falls between PAH features in starbursts, and between silicate absorption or emission features in AGN.  As a result, it accurately represents the continuum of the underlying dust emission. Continuum fluxes, redshifts, and resulting continuum luminosities $\nu$L$_{\nu}$ (15$\mu$m) for all PAH sources are given in Table 3.  This Table also includes a continuum measure at 5.5\,\um, although for PAH sources some of this "continuum" may arise from broad wings of PAH emission.   

IRS redshifts in Table 3 are determined from PAH emission features, assuming rest wavelengths of 6.2$\mu$m, 7.7$\mu$m, 8.6$\mu$m, and 11.3$\mu$m.  For the 29 sources also having optical redshifts from spectra in the Sloan Digital Sky Survey (SDSS, Gunn et al. 1998), the average difference in z between IRS and optical redshifts is 0.0013.  The SDSS optical classifications all agree with the IRS starburst classification for the sources in Table 3, except for 3 sources noted which are optically classified as Seyfert 2 and one as a narrow-line AGN. 

Spectra of PAH sources displayed in Figures 1 and in Figures 7-11 are normalized by the peak flux density at 7.7$\mu$m.  This is chosen because f$_{\nu}$(7.7\ums) is a well defined parameter for comparing with PAH sources at high redshift, where f$_{\nu}$(7.7\ums) is the brightest observed portion of the rest-frame spectrum and strongly influences the MIPS flux densities observed for faint sources \citep{wee08}.  To provide a measure of spectral slope and of dispersion among spectral shapes, the ratios of continuum flux density at rest-frame 24\,\um and 15\,\um to the peak flux density f$_{\nu}$(7.7\ums) are also given in Table 3. 
 
\subsection{Sources with Silicate features}

Figures 2 and 3 show the 16 sources from Bootes and FLS with silicate emission, and Figure 4 shows the 8 sources with silicate absorption. Spectral parameters for these silicate sources, which we classify as AGN, are listed in Tables 4 and 5.  The AGN classifications from the IRS spectra are consistent with the optical SDSS classifications (noted in Table 5) when optical spectra are available (16 of the 24 sources).  IRS redshifts for the silicate sources in Table 5 are not as accurate as for the PAH sources in Table 3, because fewer narrow spectral features are available for measure.  Only 8 silicate sources have confident redshifts from both IRS and SDSS, and the average difference in z for these is 0.0088.  

Because of the weakness of PAH features in these sources, we list in Table 4 measures (usually upper limits) only for the 6.2\,\um feature, and also give measures or limits for the [NeIII] 15.56\,\um emission line.  Table 4 also includes optical depth $\tau_{si}$ as a measure of depth for the 9.7\,\um silicate absorption feature.  This is defined as $\tau_{si}$ = ln[$f_{\nu}$(cont)/$f_{\nu}$(abs)], for $f_{\nu}$(abs) measured at the wavelength of maximum depth for absorption and $f_{\nu}$(cont) being an unabsorbed continuum at the same wavelength, extrapolated from either side of the absorption feature as in \citet{spo07}. Strength of silicate emission features is not measured because of uncertainty in defining the underlying continuum. 

The continuum flux density and luminosity at 15\,\um rest frame wavelength are given for the AGN sources in Table 5, as well as continuum flux density at 5.5\ums.  For the AGN sources, which do not have strong 7.7\,\um PAH emission, spectra are normalized using the continuum at 8$\mu$m rest frame. This wavelength is chosen because it defines a localized spectral peak of the continuum for sources with strong silicate absorption; this peak arises because of absorption features redward and blueward of the peak.  The values of f$_{\nu}$(8\ums) are given in Table 5, along with the ratios of continuum flux densities at 15\,\um and 24$\mu$m to the f$_{\nu}$(8\ums).

Of the 24 AGN in Table 4, 17 have redshifts from SDSS, and 3 additional sources without SDSS redshifts have firm IRS redshifts from well defined spectral features (strong silicate absorption or atomic emission lines).   The IRS spectra in Figures 2-4 confirm the SDSS redshifts for 8 sources; source AGN23 has an IRS spectrum that indicates z $\sim$ 1.1 compared to the SDSS redshift of 0.35.  This source is shown in Figure 2 using the IRS redshift, but because of the redshift ambiguity is not used in the average spectra discussed below.  

Sources AGN3, AGN5, AGN9 and AGN17 do not have SDSS redshifts and do not have strong silicate absorption or atomic emission features that enable a confident IRS redshift.  Of these four, sources AGN5, AGN9, and AGN17 are shown in Figures 2-4 at the estimated IRS redshift.  These estimates derive from comparisons to other spectra of known redshift by seeking the best match for rest frame spectra of these 3 sources.  Because of the uncertainty in z, however, sources AGN5, AGN9, and AGN17 are not used in deriving our average spectra.

Source AGN3 is particularly interesting.  \citet{bro06} present a redshift distribution of all type 1 QSOs in the Bootes survey field with f$_{\nu}$(24\ums) $>$ 1 mJy having optical redshifts determined from the AGN and Galaxy Evolution Survey (AGES, Cool et al. 2006).  Although individual AGES redshifts are not published, the distribution of redshifts and 24\,\um fluxes in Figure 8 of Brown et al. show that a type 1 QSO is present in Bootes having f$_{\nu}$(24\ums) $>$ 10 mJy and z = 2.4.  This source must be in our 10 mJy Bootes sample because sources were chosen from the same survey, but it is not among the sources with either SDSS redshifts or firm IRS redshifts.  

A luminous type 1 QSO should show strong silicate emission \citep{hao05}.  Of the three Bootes sources in Table 4 without firm SDSS or IRS redshifts, source 3 is the only source whose IRS spectrum is consistent with a strong silicate emission source at z = 2.4.  As seen in Figure 3, the rest frame spectrum of this source shows an increasing continuum at $\sim$ 10\,\um that is just as expected for the onset of a strong 9.7\,\um silicate emission feature.  We note also in Figure 3 the close similarity of the rest-frame spectrum of source AGN3 to that of source AGN21, a type 1 QSO with a known SDSS redshift and the most luminous source in our 10 mJy sample, as discussed below.  The MIPS flux of source AGN3 \citep{hou07} of 10.3 mJy is also in agreement with the flux of 10.5 mJy plotted by \citet{bro06}. We conclude, therefore, that source AGN3 is the QSO at z = 2.4 identified by \citep{bro06}, and we include this source in our average spectra.

\section{Discussion}

\subsection{Distribution of Redshifts and Luminosities}

Each source in Tables 2-5 is individually interesting, and many comparisons could be made among various properties because extensive multiwavelength data already exist for most of these objects.  For the present, we are not undertaking such an overall multiwavelength analysis except for noting in Tables 3 and 5 the general agreement between optical and IRS spectral classifications.  The primary result we discuss here is the relation between mid-infrared luminosity and the infrared spectral characteristics.  This result is crucial to understanding the nature of sources which are seen in surveys of the mid-infrared sky, and in using these surveys to determine evolutionary characteristics of starbursts and AGN.  

This 10 mJy sample clearly shows that the most luminous sources are those with AGN characteristics (silicate features) rather than starburst characteristics (PAH features).  In Figure 5, the redshifts and 24\,\um fluxes are compared for sources with and without measurable PAH features.  It is seen that, while the flux distributions are similar, the redshifts are generally much higher for the AGN, implying greater luminosities.  

This result is made clear in Figure 6 where luminosity distributions are shown.  Luminosities are compared using rest-frame $\nu$L$_{\nu}$(15$\mu$m) in  ergs s$^{-1}$.  (Log $\nu$L$_{\nu}$(15$\mu$m)(\ldot) = log $\nu$L$_{\nu}$(15$\mu$m)(ergs s$^{-1}$) - 33.59.)  As mentioned previously, a wavelength of 15\,\um is used for comparison of intrinsic luminosities because the 15\,\um continuum is a pure dust continuum, without contamination by PAH or emission line spectral features and is between the 10\,\um and 17\,\um silicate features.  Being an intermediate mid-infrared wavelength, it is not as dominated by luminosity from the hot dust associated with AGN or the cool dust associated with a starburst as would be a shorter or longer wavelength.

In Figure 6, luminosities $\nu$L$_{\nu}$(15$\mu$m) are compared to the EW of the 6.2\,\um PAH feature.  This Figure shows our most important results.  It shows the dramatic range in dust continuum luminosities that is covered by the sample, a factor of 2.5 x 10$^{4}$ in luminosity, a range that encompasses virtually all extragalactic sources observed so far with $Spitzer$, regardless of selection criterion. The faintest source is source SB11 in Tables 2 and 3, a blue compact dwarf in Bootes with log $\nu$L$_{\nu}$ (15$\mu$m) = 41.85 (ergs s$^{-1}$); the most luminous sources are sources AGN11 and AGN21 in Tables 4 and 5, with AGN11 being an absorbed silicate source and AGN21 an emission silicate source, with log $\nu$L$_{\nu}$ (15$\mu$m) = 46.2. (Source AGN3 in Table 5, the QSO at z = 2.4 discussed in section 2.3, does not have a measurement of $\nu$L$_{\nu}$ (15$\mu$m) because the observed rest-frame spectrum does not extend to 15\,\um. It certainly is among the sources with log $\nu$L$_{\nu}$(15$\mu$m) $>$ 45.0 (ergs s$^{-1}$) so is included in the average for this luminosity bin discussed below in section 3.2.)

Figure 6 shows a well defined gap in the distribution of PAH strength at 0.4\um $<$ EW(6.2\ums) $<$ 0.5\ums.  21 of the 60 sources show EW(6.2\ums) $>$ 0.47\ums, and the remaining sources all have EW(6.2\ums) $<$ 0.37\ums.  We also note that the "pure" starbursts in \citet{bra06} show a lower limit for EW(6.2\ums) at about this value; 21 of 22 Brandl et al. starbursts have EW(6.2\ums) $>$ 0.4\ums.  We interpret these empirical results for both our 10 mJy sample and the Brandl et al. sample to mean that sources with EW(6.2\ums) $>$ 0.4\um are pure starbursts.  Sources in which the strength of the PAH feature is diluted by additional mid-infrared continuum arising from an AGN would show EW(6.2\ums) $<$ 0.4\ums, and we consider such sources as composite starburst+AGN.  Sources with the smallest values of EW(6.2\ums) are dominated by the AGN component.  These interpretations lead to the classifications shown in Figure 6. 

For starbursts with EW(6.2\ums)$>$ 0.4\ums, the median log $\nu$L$_{\nu}$ (15$\mu$m) = 43.1.  For composite sources with 0.1\um $<$ EW(6.2\ums)$<$ 0.4\ums, the median log $\nu$L$_{\nu}$ (15$\mu$m) = 44.0. For AGN sources with EW(6.2\ums)$<$ 0.1\ums, the median log $\nu$L$_{\nu}$ (15$\mu$m) = 45.0.

\subsection{Average Spectra for Different Luminosities}

The change in the nature of mid-infrared spectra as a function of luminosity is also clearly shown in average spectra binned with luminosity.  We consider 4 bins of luminosity: log $\nu$L$_{\nu}$(15$\mu$m) $>$ 45.0 (ergs s$^{-1}$), 45.0 $>$ log $\nu$L$_{\nu}$(15$\mu$m) $>$ 44.0, 44.0 $>$ log $\nu$L$_{\nu}$(15$\mu$m) $>$ 43.0, and 43.0 $>$ log $\nu$L$_{\nu}$(15$\mu$m) $>$ 42.0.  The normalized average spectra within these bins are shown in Figure 7 and tabulated in Table 6.  The most important conclusion from Figure 7 is the progressive increase in PAH strength as luminosity $\nu$L$_{\nu}$(15$\mu$m) decreases.  

While these average spectra can be used as empirical spectral templates for different luminosities, the dispersion among spectra illustrates the uncertainties which arise when adopting a specific template. To allow an estimate of this dispersion within a luminosity bin, Figures 8-11 show the continuum flux densities at 15\,\um and 24\,\um relative to the normalizing peak at 7.7\um (starbursts) or 8\um (AGN) for all of the individual sources which enter an average.  The Figures also show the most extreme sources in each bin.  

All sources do not cover all rest frame wavelengths because of varying redshifts; for example, only 3 of the most luminous sources are at sufficiently low redshift to allow a measure of rest frame f$_{\nu}$(24\ums).  Average spectra are determined within different wavelength ranges using only the sources whose rest-frame spectra include those wavelength ranges.

\subsection{Star Formation Rate Indicators for Starbursts}

We use the classification "starburst galaxies", or "starbursts", to mean galaxies whose spectra and luminosity are dominated by the consequences of on-going star formation. A fundamental measurement of importance for starbursts is the measure of star formation rate (SFR).  Mid-infrared spectra of starbursts such as those in Figure 1 contain data for three indicators that give independent measures of SFR.  These are the luminosity of the PAH emission features \citep[e.g. ][]{for04,bra06,hou07}, the luminosity of the dust continuum \citep[e.g. ][]{ken98,cal07}, and the luminosity of the Neon emission lines \citep{ho07}.  These three indicators measure in different ways the luminosity arising from the young stars of the starburst.  The PAH emission is excited by photons penetrating the photodissociation region (PDR) at the boundary between the HII region and the surrounding molecular cloud; the dust continuum arises from dust intermixed in the HII region and heated by the stars; and the emission lines arise within the HII region from ionizing photons arising in the young stars. 

If the geometry of the star-forming environment, the spectral distribution of stellar radiation, the temperature distribution and nature of the dust, and the gas to dust ratio were the same for all starbursts, any one of these parameters should give the same result for SFR.  Of course, all starbursts are not the same among these many characteristics, so different indicators of SFR give different results depending on these characteristics.  Although we do not attempt here an evaluation of the relative merits of various estimates of SFR, we can use the data for our 10 mJy sample of starbursts to estimate the dispersion that arises among these three different methods for measuring SFR.  

For this comparison, we utilize the 7.7\,\um PAH feature, the [NeIII] 15.56\,\um feature, and the strength of the dust continuum at 24\,\um. The flux $\nu$f$_{\nu}$(7.7$\mu$m) at the peak of the PAH feature measures the luminosity in the photodissociation region, the flux of the Neon line measures the ionizing luminosity within the HII region, and the flux density of the continuum measures the emission from warm dust. The measures used for comparison are given in Tables 2 and 3 and illustrated in Figures 12 and 13. Parameters are compared to EW(6.2\,\um) which determines if a source spectrum arises strictly from a starburst without AGN contamination, as discussed in section 3.1.

In Figure 12, the distribution of the ratio $\nu$f$_{\nu}$(7.7$\mu$m) to f([NeIII]) is shown for the sources in Tables 2 and 3 with PAH features.  This is a measure of luminosity arising in the PDR compared to that arising in the HII region.  Including the limits which are shown, there is a weak trend in Figure 12 for PAH EW to increase as the ratio of PAH to [NeIII] increases.  This trend is as expected if the weak PAH sources contain an AGN contribution to the [NeIII] luminosity.  The comparison of SFRs is intended to apply only for sources which are pure starbursts, without an AGN contribution. These pure starbursts are taken as defined in Figure 6, based on a criterion that EW(6.2\ums) $>$ 0.4\ums. 

For these pure starbursts, the median and dispersion in the PAH to [NeIII] ratio in Figure 12 are log [$\nu$f$_{\nu}$(7.7$\mu$m)/f([NeIII])] = 2.8$\pm$0.3.  This result indicates that we should expect a dispersion by a factor of $\pm$ 2.0 between SFR estimates from PAH compared to those from [NeIII], even for pure starbursts. 

In Figure 13, the distribution of the ratio f$_{\nu}$(24$\mu$m) to f$_{\nu}$(7.7$\mu$m) is shown for the sources in Tables 2 and 3 with PAH features. This is a measure of luminosity arising from warm dust within both the HII region and PDR compared to luminosity arising in the PDR.  There is a weak trend for this ratio to change with EW(6.2\ums).  This trend is expected, simply because EW(6.2\ums) is also a measure of PAH strength compared to the continuum.  For the pure starbursts with EW(6.2\ums) $>$ 0.4\ums, the median and dispersion of the ratio are log[f$_{\nu}$(24$\mu$m)/f$_{\nu}$(7.7$\mu$m)] = 0.3$\pm$0.3; the dispersion is the same as for the dispersion in PAH to [NeIII].  This result indicates that either PAH measures or dust continuum measures would give the same value for SFR to within a factor of two.

This analysis considers only the empirical scatter among the three mid-infrared parameters which can be used for SFR measures, but it does not explain the sources of this scatter.  Given the many assumptions which enter the transformation from an observed spectral parameter to a SFR, the utility of our empirical result is primarily to conclude that it is possible to estimate the SFR to a factor of $\sim$ 2 using several indicators within a mid-infrared spectrum, but more detailed assumptions regarding the nature of the starburst would be necessary to improve such estimates.

\subsection{Predicted MIPS Fluxes for Different Average Spectra}

Average spectra are important for adopting spectral templates as a function of luminosity for use in modeling source counts \citep[e.g. ][]{cha04,lag04,lef05,cap07}.  Such modeling is the fundamental source of conclusions regarding the evolution of extragalactic infrared sources, and different assumptions regarding templates give different answers.  A utility of the average spectra in Figures 7 through 11 and Table 6 is that these comprise an empirically derived set of templates and dispersions among the templates which arise directly from a flux limited sample of sources selected only as infrared sources.  These templates can be used with assumptions for luminosity functions and evolution of sources to predict source counts at mid-infrared wavelengths, although we do not at present undertake such predictions. 

For now, we use the four average spectra and corresponding luminosities only to illustrate the observed MIPS f$_{\nu}$(24\ums) which would arise at different redshifts from the templates for the various luminosity bins shown in Figures 7 through 11. The determination of MIPS f$_{\nu}$(24\ums) is made using the synthetic photometry tool in SMART and relating flux to luminosity with a cosmology having H$_0$ = 71 \kmsMpc, $\Omega_{M}$=0.27 and $\Omega_{\Lambda}$=0.73.  Results are in Figure 14. 

A particularly important implication of the results in Figure 14 is that the sources at the highest redshifts which are detected in MIPS surveys with f$_{\nu}$(24\ums) $\ga$ 1 mJy should only be the luminous AGN without strong PAH features, sources like those in Figure 11 having log $\nu$L$_{\nu}$(15$\mu$m) $>$ 45.0. For sources in this luminosity bin, the observed MIPS f$_{\nu}$(24\ums) remains above 1 mJy to a redshift of 3.  By contrast, sources in luminosity bins whose average spectra show strong PAH features (log $\nu$L$_{\nu}$(15$\mu$m) $<$ 44.0) would drop to f$_{\nu}$(24\ums) $<$ 0.1 mJy for z $>$ 1.  

In fact, however, this prediction is proven incorrect by the large numbers of PAH-dominated sources at z $\sim$ 2 and f$_{\nu}$(24\ums) $\ga$ 1 mJy which have been discovered by $Spitzer$ \citep[e.g. ][]{yan07,far08,pop08}.  Explaining such sources requires luminosity evolution for starbursts such that PAH dominated spectra can be found in sources with log $\nu$L$_{\nu}$(15$\mu$m) $>$ 45.0.  By considering only the most luminous starbursts discovered so far using a PAH luminosity indicator, this evolution has been quantified to scale as (1+z)$^{2.5}$ for the most luminous starbursts \citep{wee08}.  Applying evolution by this factor would raise the curve for log $\nu$L$_{\nu}$(15$\mu$m) = 44.0 in Figure 14 to fluxes more than a factor of 10 brighter than shown and would indeed predict PAH sources to be found at z $\sim$ 2 with f$_{\nu}$(24\ums) $\sim$ 1 mJy.

This result is illustrated in Figure 14 by taking the most extreme spectra in Figure 10 for 44.0 $<$ log $\nu$L$_{\nu}$(15$\mu$m) $<$ 45.0 and asssigning luminosities log $\nu$L$_{\nu}$(15$\mu$m) = 45.0 to spectra with those extreme shapes.  The lower spectrum, having strong PAH features, would reach f$_{\nu}$(24\ums) = 1 mJy at z = 2 using luminosity log $\nu$L$_{\nu}$(15$\mu$m) = 45.0, although the actual luminosity of this source is log $\nu$L$_{\nu}$(15$\mu$m) = 44.3.

 \section{Summary and Conclusions}

Spectra are presented and discussed for a flux-limited sample of 60 galaxies with f$_{\nu}$(24\ums) $>$ 10\,mJy taken from within the $Spitzer$ First Look Survey and the NOAO Deep Wide-Field Survey region in Bootes.  36 sources are characterised as starbursts based on the presence of PAH features, and 24 sources as AGN based on the presence of silicate emission or absorption features.  Sources have 0.01 $<$ z $<$ 2.4.  

The distribution of luminosities and the average mid-infrared spectra of sources as a function of luminosity are determined defined by the continuum luminosity at 15$\mu$m. Lower luminosity sources are dominated by PAH emission features, and higher luminosity sources are dominated by silicate absorption or emission.  Source luminosity $\nu$L$_{\nu}$(15$\mu$m) increases as the equivalent width of the 6.2\um PAH feature decreases.  For sources with EW(6.2\ums)$>$ 0.4\ums, the median log $\nu$L$_{\nu}$(15$\mu$m) = 43.1 (ergs s$^{-1}$).  For sources with 0.1\ums $<$ EW(6.2\ums)$<$ 0.4\ums, the median log $\nu$L$_{\nu}$(15$\mu$m) = 44.0. For sources with EW(6.2\ums)$<$ 0.1\ums, the median log $\nu$L$_{\nu}$(15$\mu$m) = 45.0. 

Average spectra are used to predict the $Spitzer$ MIPS flux densities f$_{\nu}$(24\ums) that would be observed as a function of redshift for sources of different luminosities. Results show that without significant luminosity evolution, the sources that should be seen with f$_{\nu}$(24\ums) $\ga$ 1 mJy and z $\sim$ 2 would have silicate absorption or emission features and no PAH features.  Because PAH sources have been discovered by $Spitzer$ with f$_{\nu}$(24\ums) $\ga$ 1 mJy and z $\sim$ 2, substantial luminosity evolution is required for sources with PAH-dominated spectra.

For the pure starbursts, dispersions among PAH strength, [NeIII] emission line strength, and dust continuum strength are illustrated to estimate the resulting dispersions among star formation rates that would be derived from these three independent indicators of SFR.  It is found that the dispersions indicate an uncertainty of about a factor of two in deriving the SFR from these different indicators.

\acknowledgments
We thank P. Hall for help in improving our IRS spectral analysis with SMART.  This work is based primarily on observations made with the
Spitzer Space Telescope, which is operated by the Jet Propulsion
Laboratory, California Institute of Technology under NASA contract
1407. Support for this work by the IRS GTO team at Cornell University was provided by NASA through Contract
Number 1257184 issued by JPL/Caltech. This research has made use of the NASA/IPAC Extragalactic Database (NED) which is operated by the Jet Propulsion Laboratory, California Institute of Technology, under contract with the National Aeronautics and Space Administration.

\clearpage
\pagestyle{empty}

\begin{deluxetable}{lccccc} 
\tablecolumns{6}
\tabletypesize{\footnotesize}

\tablewidth{0pc}
\tablecaption{The $>$10 mJy Extragalactic Point Source Sample in the FLS}
\tablehead{
  \colhead{Source}& \colhead{Source Name\tablenotemark{a}}& \colhead{AOR\tablenotemark{b}}& \colhead{program}& \colhead{f$_{\nu}$(24$\mu$m)\tablenotemark{c}}& \colhead{IRS class\tablenotemark{d}} \\
  \colhead{}& \colhead{}& \colhead{}& \colhead{}& \colhead{mJy}& \colhead{}
}
\startdata

1 & SST24 J171744.12+583848.9   &21651200  &40038   &13.9(14.2) & PAH, 1C\\
2 & SST24 J171352.41+584201.2  &21649152 & 40038 & 23.9(23.5)   & Si em, 1A\\

3 & SST24 J171207.43+584754.4   & 21651456  &40038     &13.3(13.6)  & Si abs, 1A\\

4 & SST24 J171335.15+584756.1   & 21649664 &40038   &23.7(22.8)    & Si abs, 1A\\

5 & SST24 J171150.21+590041.7  &21652224  &40038    &10.9(10.6)  & PAH, 1A\\

6 & SST24 J172400.61+590228.4  &14126848  &20128    &11.6(14.3)  & PAH, 1C\\

7 & SST24 J171852.71+591432.0    &24187904  & 40539   &13.9(14.5)  & Si abs, 2B\\

8 & SST24 J171607.21+591456.3  &21648640 &40038   &34.6(35.3) & PAH, 1C\\

9 & SST24 J171542.03+591657.6  &14127616  &20128    &26.3(28.1)  & PAH, 1C\\

10 & SST24 J171021.76+591854.7    &21650688  &40038   &14.7(14.5)    &Si abs, 1A\\ 

11 & SST24 J171641.09+591857.2    & 21649920 &40038   &20.9(21.3)    & PAH, 1C\\

12 & SST24 J172825.80+592653.3  &21652992  &40038    &10.2(10.3)  & no features, 1A\\

13 & SST24 J172611.96+592851.8   &21650432  &40038    &16.2(17.0)  & PAH, 1C\\

14 & SST24 J171839.73+593359.6   &21648896  &40038  &13.2(13.8)    &Si em, 1A\\

15 & SST24 J171916.62+593449.6  &14126336  &20128    &17.4(19.2)  & PAH, 1B\\ 

16 & SST24 J171302.37+593611.0  &21651968  &40038     &11.8(11.4)  &Si abs, 3A\\

17 & SST24 J171902.29+593715.9 &21648384  &40038    & 26.9(26.9) &Si em, 1A\\ 
 
18 & SST24 J172704.67+593736.6    &21649408  &40038   &23.2(23.1)   & Si em, 1A\\

19 & SST24 J171437.42+595648.0  &14127360  &20128    &11.6(12.0)  & PAH, 1B\\
 
20 & SST24 J171530.75+600216.4   &21651712  &40038    &11.6(11.8)  & Si abs. and em., 1A\\

21 & SST24 J172643.84+600238.7    &21650176  &40038    &18.0(18.0)  & PAH, 1B\\
 
22 & SST24 J172123.19+601214.6  &14016768   &20083    & 13.3(13.2) & uncertain, 1A\\
 
23 & SST24 J171634.03+601443.6  &21652480  &40038    &10.8(11.3)  & PAH, 1C\\

24 & SST24 J170829.79+601856.4   &21652736  &40038    &14.3(15.3)  & weak PAH, 1A\\

25 & SST24 J171313.96+603146.6 &21653248  &40038     &10.5(10.2)  & no features, 1A\\

\enddata

\tablenotetext{a}{SST24 source name derives from discovery with the
  MIPS 24$\mu$m images; coordinates listed are J2000 24$\mu$m positions with typical 3\,$\sigma$ uncertainty of $\pm$ 1.2\arcsec. }
\tablenotetext{b}{AOR number is the observation number in the $Spitzer$ archive with the program number which is given.}
\tablenotetext{c}{Values of f$_{\nu}$(24$\mu$m) are for an unresolved point source in the FLS measured from MIPS 24\,\um images. Values in parentheses are the f$_{\nu}$(24$\mu$m) measured independently using synthetic photometry of the extracted IRS spectra; the mean difference between IRS and MIPS values indicates that uncertainties in photometry are less than $\pm$ 5\%.}

\tablenotetext{d}{Classification of IRS spectrum, whether showing PAH emission, absorption by the 9.7$\mu$m silicate feature, or emission by the 9.7$\mu$m silicate feature; numerical value gives classification according to scheme of \citet{spo07}. More data for PAH sources are in Tables 2 and 3 and Figure 1; more data for silicate sources are in Tables 4 and 5 and Figures 2, 3, and 4.}

\end{deluxetable}

\begin{deluxetable}{lcccccccc} 
\rotate
\tablecolumns{9}
\tabletypesize{\footnotesize}

\tablewidth{0pc}
\tablecaption{PAH and [NeIII] Fluxes for Starbursts in FLS and Bootes} 
\tablehead{
  \colhead{Source} & \colhead{Coordinate\tablenotemark{a}}& \colhead{f$_{\nu}$(7.7\ums)\tablenotemark{b}}&\colhead{f(6.2\ums)\tablenotemark{c}} & \colhead{EW(6.2\ums)} &\colhead{f(11.3\ums)\tablenotemark{d}} & \colhead{EW(11.3\ums)}&  \colhead{[NeIII]\tablenotemark{e}} & \colhead{EW(15.56\ums)}\\ 
  \colhead{}&  \colhead{}  &\colhead{mJy} &\colhead{}& \colhead{\ums}& \colhead{}&  \colhead{\ums }& \colhead{15.56\ums} &  \colhead{\ums }}

\startdata

SB1 &142646.63+322125.6   &9.2  & 50   & 0.51$\pm$0.05  & 30   & 0.95$\pm$0.06  & 8.8 & 0.12$\pm$0.03 \\
SB2 &  142629.15+322906.7   & 6.9 &42 &0.32$\pm$0.03 &48 &0.66$\pm$0.04 & $<$2 &$<$0.03 \\
SB3 &  142623.91+324436.0  & 10.3  &43 &0.21$\pm$0.02 &26 &0.35$\pm$0.05 &$<$5 &$<$0.06 \\
SB4 &  143115.23+324606.2   & 9.8 &68  &0.57$\pm$0.04 &67 &0.65$\pm$0.04 &12.5 &0.17$\pm$0.2 \\
SB5 &  143125.46+331349.8    & 15.1  &152 &0.75$\pm$0.03  &106 &0.74$\pm$0.03 &45 &0.43$\pm$0.05 \\
SB6 &  142659.15+333305.1   &16.4  &85 &0.54$\pm$0.02 &64 &0.60$\pm$0.05 &$<$4 &$<$0.02 \\ 
SB7 &  143156.25+333833.3  &27.8 &186 & 0.48$\pm$0.02 &186 &0.76$\pm$0.04 &10.0 &0.03$\pm$0.01 \\
SB8 &   142543.59+334527.6   &12.3 &76  &0.52$\pm$0.3  &76 &0.34$\pm$0.02 &8.6 &0.03$\pm$0.01 \\
SB9 &  143632.01+335230.7  & 9.4 &80  & 0.83$\pm$0.05  &54 &0.73$\pm$0.03 &$<$3 &$<$0.03 \\
SB10 & 142552.71+340240.2   &7.9 &13 &0.08$\pm$0.02  &22 &0.34$\pm$0.08 &$<$8 &$<$0.06 \\
SB11 & 143232.52+340625.2  &2.2  &15.2  &0.22$\pm$0.04  &15.5 &0.31$\pm$0.03 &6.5 &0.10$\pm$0.02 \\
SB12 & 143120.00+343804.2  &2.7 &18  & 0.33$\pm$0.09  &15.7 &0.22$\pm$0.2 &25.2 &0.22$\pm$0.03 \\
SB13 & 143631.98+343829.3  &18.9  &30  &0.04$\pm$0.01  &14.3 &0.03$\pm$0.01 &$<$5 &$<0.02$ \\
SB14 &  143126.81+344517.9  &13.7 &94  & 0.51$\pm$0.04  &76 &0.75$\pm$0.05 &$<$2 &$<0.02$ \\
SB15 & 142554.57+344603.2 &38.6  &286  &0.55$\pm$0.02  &218 &0.75$\pm$0.05 &12.5 &0.05$\pm$0.01 \\
SB16 & 142504.04+345013.7 & 16.6 &112  & 0.59$\pm$0.02  &87 &0.82$\pm$0.04 &4.4 &0.05$\pm$0.01 \\
SB17 &  143641.26+345824.4  &  21.8  & 145 & 0.61$\pm$0.03  &168 &0.83$\pm$0.03 &11 &0.08$\pm$0.01 \\
SB18  & 143053.70+345836.7  &9.2  &36  & 0.08$\pm$0.01 &40 &0.12$\pm$0.02 &$<$9 &$<$0.02\\
SB19 & 143239.59+350151.5  &9.2 &42  & 0.30$\pm$0.03  &37 &0.54$\pm$0.03 &8 &0.11$\pm$0.02 \\
SB20 & 142417.44+342046.7   &89 &1200  &0.64$\pm$0.02   &1360 &0.79$\pm$0.02 &52 &0.05$\pm$0.01 \\
SB21 & 143046.32+351313.6  &  9.7  &63& 0.31$\pm$0.03  &48 &0.37$\pm$0.02 &4.4 &0.04$\pm$0.01 \\

SB22 &  143039.27+352351.0  & 14.4 &87  &0.47$\pm$0.3  &71 &0.53$\pm$0.02 &$<$10 &$<$0.05\\
SB23 & 143119.79+353418.1 & 39  &306  & 0.67$\pm$0.03 &240 &0.79$\pm$0.02 &15 &0.08$\pm$0.02 \\

SB24 & 143121.15+353722.0  &52  &340  &0.51$\pm$0.01 &200 &0.70$\pm$0.02 &14 &0.04$\pm$0.01 \\

SB25 & 171744.12+583848.9  & 2.8  &25 &0.64$\pm$0.05 &12 &0.32$\pm$0.03 &3.8 &0.05$\pm$0.02  \\

SB26 & 171150.21+590041.7  & 1.3 &$<$9  &$<$0.08   &8.8  &0.18$\pm$0.02 &$<$5 &$<$0.08  \\

SB27  & 172400.61+590228.4 &6.3 &38 &0.51$\pm$0.04 & 29  &0.47$\pm$0.04  &$<$3  &$<$0.04  \\

SB28 & 171607.21+591456.3 &15.5\tablenotemark{f}   &111 &0.61$\pm$0.03 &101 & 0.70$\pm$0.04&7.0  &0.06$\pm$0.01   \\

SB29  &171542.03+591657.6  &11.7    &98  & 0.54$\pm$0.03  &47 &0.49$\pm$0.04  &13  & 0.10$\pm$0.03 \\

SB30 & 171641.09+591857.2    & 10.5\tablenotemark{f}   & 81  &0.69$\pm$0.04  &61  &0.68$\pm$0.04  &8.3 &0.11$\pm$0.02  \\

SB31 & 172611.96+592851.8   & 13.0\tablenotemark{f}  & 91  & 0.55$\pm$0.03  &78  & 0.71$\pm$0.04  & 4.7 & 0.05$\pm$0.01 \\

SB32  & 171916.62+593449.6  & 7.6   &35  &0.28$\pm$0.2  & 17 & 0.27$\pm$0.03  &14   & 0.14$\pm$0.05 \\
SB33 & 171437.42+595648.0     &4.3   &18  &0.32$\pm$0.06   &19  &0.35$\pm$0.04   & 7.2 & 0.14$\pm$0.05 \\

SB34 & 172643.84+600238.7  & 8.3 &50  &0.36$\pm$0.02  & 44 &0.44$\pm$0.03  &$<$6   &$<$0.06 \\
SB35 & 171634.03+601443.6    &5.1\tablenotemark{f} &35  &0.74$\pm$0.08  &30  & 0.60$\pm$0.06  & 14 & 0.30$\pm$0.02 \\

SB36 & 170829.79+601856.4      &5.1  &5.0  &0.03$\pm$0.02  &17  &0.13$\pm$0.03   &$<$4  &$<$0.03  \\

\enddata

\tablenotetext{a}{Coordinates identify sources from Table 1 for FLS sources or from Table 1 in \citet{hou07} for Bootes sources.}
\tablenotetext{b}{Flux density at peak of 7.7$\mu$m PAH feature used for normalizing spectra shown in Figure 1; values taken from \citet{hou07} for Bootes sources.}
\tablenotetext{c}{Total flux of 6.2\,\um PAH emission feature in units of 10$^{-22}$W cm$^{-2}$, fit with single gaussian on a linear continuum within rest wavelength range 5.5\,\um to 6.9\,\um; uncertainties deriving from noise in the feature and from flux calibration are typically $\pm$ 10\%. Values are from new spectral extractions for FLS sources in Figure 1 and taken from \citet{hou07} for Bootes sources, except that new extractions were used for Bootes sources from the $Spitzer$ public archive in program 20113 (H. Dole, P.I.). Equivalent widths (EW) are in the rest frame.}
\tablenotetext{d}{Total flux of 11.3\,\um PAH emission feature in units of 10$^{-22}$W cm$^{-2}$, fit with single gaussian on a linear continuum within rest wavelength range 10.4\,\um to 12.2\,\um; uncertainties deriving from noise in the feature and from flux calibration are typically $\pm$ 10\%. Values are from new spectral extractions for FLS sources in Figure 1 and taken from \citet{hou07} for Bootes sources, except that new extractions were used for Bootes sources from the $Spitzer$ public archive in program 20113.  Equivalent widths (EW) are in the rest frame. }
\tablenotetext{e}{Total flux of [NeIII] emission feature in units of 10$^{-22}$W cm$^{-2}$, fit with single gaussian on a linear continuum within rest wavelength range 14.9\,\um to 16.2\,\um; uncertainties deriving from noise in the feature and from flux calibration are typically $\pm$ 10\%. Values are from new spectral extractions for FLS sources in Figure 1 and taken from \citet{hou07} for Bootes sources, except that new extractions were used for Bootes sources from the $Spitzer$ public archive in program 20113.  Equivalent widths (EW) are in the rest frame.}
\tablenotetext{f}{Differs from incorrect value tabulated in \citet{wee08}.}

\end{deluxetable}

\clearpage
\pagestyle{empty}

\begin{deluxetable}{lcclccccc} 
\rotate
\tablecolumns{9}
\tabletypesize{\footnotesize}

\tablewidth{0pc}
\tablecaption{Redshifts and Continuum Luminosities for Starbursts in FLS and Bootes} 

\tablehead
{

\colhead{Source} & \colhead{Coordinate\tablenotemark{a}} & \colhead{z(IRS)} & \colhead {z(SDSS)\tablenotemark{b}}  & \colhead{f$_{\nu}$(5.5\ums)\tablenotemark{c}}& \colhead{f$_{\nu}$(15\ums)\tablenotemark{d}} &\colhead{log[$\nu$L$_{\nu}$(15 \ums)]\tablenotemark{e}} &\colhead{f$_{\nu}$(15\ums)/f$_{\nu}$(7.7\ums)}  & \colhead{f$_{\nu}$(24\ums)/f$_{\nu}$(7.7\ums)}\\ 

\colhead{} & \colhead {} & \colhead{} & \colhead{} &\colhead {mJy} &\colhead {mJy} &\colhead{}  &\colhead{} &\colhead{}
}

\startdata
SB1 & 142646.63+322125.6 & 0.2683 & \nodata& 1.5 &5.8 & 44.30 & 0.41 & 1.5 \\ 
SB2 &  142629.15+322906.7 & 0.1028 & 0.1008 &0.89  &4.4 &43.32  &0.58 & 2.5  \\ 
SB3 &   142623.91+324436.0 & 0.1760 & \nodata &3.5  &8.7 &44.10  &0.82 & 2.3   \\
SB4 &   143115.23+324606.2 & 0.0244 & 0.0227 &0.82  &6.6 &42.23 &0.57 &2.4 \\ 
SB5 &   143125.46+331349.8 & 0.0233 & 0.0226 &1.6  &8.1 &42.28  &0.48 &2.7 \\ 

SB6 &   142659.15+333305.1 & 0.1548 & \nodata &1.07  &6.5 &43.86  &0.41 &1.69 \\
SB7 &   143156.25+333833.3 & 0.0329 & \nodata & 3.0 &20.3 &42.98  & 0.73 & 1.8 \\ 

SB8 &   142543.59+334527.6 & 0.0715 & 0.0717 &1.14 &15.8 &43.55 &1.25 &4.3 \\ 
SB9 &   143632.01+335230.7 &0.0883 & 0.0866& &5.3 &43.26  & 1.23  &4.4 \\ 
SB10 &    142552.71+340240.2 &0.5642 & \nodata&1.59 &10.7 &45.24 & 1.42 &\nodata \\ 
SB11 & 143232.52+340625.2 &0.0444  &0.0423(Sy2)  &0.52  &2.6  & 42.35 &0.93  &3.6  \\
SB12 &  143120.00+343804.2&0.0156 & 0.0146 &0.50  &6.8 &41.85 &2.7  &10.7  \\
SB13 & 143631.98+343829.3 & 0.354& \nodata & 8.2 &35.7  &45.34  &1.9 &5.4  \\
SB14 &   143126.81+344517.9& 0.0835 & 0.0827 &1.07  &3.8 &43.07  & 0.28 &1.00 \\ 
SB15 &   142554.57+344603.2&0.0350 & 0.0344 &3.0  &15.5  &42.92  &0.37 & 1.51\\ 
SB16 &   142504.04+345013.7& 0.0783 & 0.0768 &1.04 &5.7 &43.19 & 0.33 & 0.86 \\ 
SB17 &   143641.26+345824.4& 0.0290 & 0.0302 &1.78 &9.5  &42.54 &0.42  &1.05 \\  
SB18  &  143053.70+345836.7 &0.0853  &0.084(Sy2)  &1.9  &33.5  &44.04  &3.2 & 9.9\\ 
SB19 &  143239.59+350151.5&0.2371 &0.2357 &1.66 &5.9 &44.20  &  0.64 &2.4 \\ 
SB20 &  142417.44+342046.7& 0.0289 & 0.0285 &1.03  &70.0 &43.39  &0.38  &1.22 \\ 
SB21 &  143046.32+351313.6& 0.0838 & 0.0828(Sy2) &1.80 &7.9 &43.39 &0.82 &2.1 \\ 
SB22 &  143039.27+352351.0&0.0885 & 0.0872 &1.50  & 12.2&43.63  & 0.85 &3.4 \\ 
SB23 & 143119.79+353418.1&0.0333&0.0347 &3.0 &13.1  &42.80  &0.33  &0.96 \\ 

SB24 &  143121.15+353722.0 & 0.0347 & 0.0347 &4.4  &27.2 &43.15 & 0.51 &1.01\\ 


SB25 &171744.12+583848.9   &0.0603 &0.0655 & 0.35  & 4.6  &42.87  &1.64 &5.8  \\

SB26 &171150.21+590041.7  &0.0605 &0.0605  &0.29  &3.7  &42.78  & 3.7 &11.6  \\

SB27 &172400.61+590228.4  & 0.1783 &0.1780 & 0.71  & 6.7 &43.99  &  1.01  & 3.6 \\

SB28 &171607.21+591456.3  &0.0544 &0.0542 &1.32 &8.8 & 43.06  &0.57 & 2.6 \\  

SB29 &171542.03+591657.6  &0.1197  &0.1158  &1.5  &10.6 & 43.84 & 0.90  & 3.6 \\

SB30 &171641.09+591857.2    &0.0570   &0.0559    &1.09  &6.1 &42.94 &0.56  &2.3 \\  

SB31 &172611.96+592851.8   &0.0729   &0.0728(NLAGN)   &0.91  &6.4  &43.18 &0.47    &1.56  \\

SB32 &171916.62+593449.6  & 0.1692  &0.1657  &1.07  &7.9 &44.02  &1.00   & 4.8  \\

SB33 &171437.42+595648.0  &0.1963   &0.1962   &0.26  & 4.8 &43.94 & 1.07  & 4.9 \\ 

SB34 &172643.84+600238.7    &0.0789   &0.0799  &1.43  &7.2 &43.30 & 0.85   & 2.6 \\ 
 
SB35 &171634.03+601443.6  &0.1057  &0.1079 &0.36  &3.9 &43.29 &0.76  & 2.9  \\

SB36 &170829.79+601856.4  &0.145  &\nodata  &2.1  &7.7  &43.87  &1.55   &3.8  \\  

\enddata
\tablenotetext{a}{Coordinates identify sources from Table 1 for FLS sources or from Table 1 in \citet{hou07} for Bootes sources.}
\tablenotetext{b}{Optical redshift from SDSS as given in National Extragalactic Database (NED) archive. All SDSS spectral classifications are as starburst except those noted in parentheses.}
\tablenotetext{c}{Continuum flux density at 5.5\,\um, measured as median flux between 5.0\,\um and 6.0\,\um; uncertainties are typically $\pm$ 5\%, based on uncertainties in flux calibration.}
\tablenotetext{d}{Continuum flux density at 15\,\um, measured as median flux between 14.8\,\um and 15.2\,\um; uncertainties are typically $\pm$ 5\%, based on uncertainties in flux calibration.}
\tablenotetext{e}{Rest frame luminosity $\nu$L$_{\nu}$(15$\mu$m) in ergs s$^{-1}$ determined using f$_{\nu}$(15$\mu$m) and luminosity distances from E.L. Wright, http://www.astro.ucla.edu/~wright/CosmoCalc.html, for H$_0$ = 71 \kmsMpc, $\Omega_{M}$=0.27 and $\Omega_{\Lambda}$=0.73.  Log [$\nu$L$_{\nu}$(15$\mu$m)(\ldot)] = log [$\nu$L$_{\nu}$(15$\mu$m)(ergs s$^{-1}$)] - 33.59.} 

\end{deluxetable}

\clearpage
\pagestyle{empty}

\begin{deluxetable}{lcccccc} 

\tablecolumns{7}
\tabletypesize{\footnotesize}

\tablewidth{0pc}
\tablecaption{Spectral Features for AGN in FLS and Bootes}  
\tablehead{
  \colhead{Source} & \colhead{Coordinate\tablenotemark{a}} &\colhead{f(6.2\ums)\tablenotemark{b}} & \colhead{EW(6.2\ums)} &  \colhead{[NeIII]\tablenotemark{c}} & \colhead{EW(15.56\ums)} & \colhead{$\tau_{si}$\tablenotemark{d}}\\ 
  \colhead{} &\colhead{}& \colhead{} & \colhead{\ums}& \colhead{15.56\ums} &\colhead{\ums }&  \colhead{ }
}

\startdata


AGN1 &  143156.40+325138.1 & $<$0.9   & $<$0.01 &5.1  &0.05$\pm$0.01   & emit   \\ 
AGN2 & 143205.63+325835.2 &$<$6  &$<$0.04   &$<$4  &$<$0.03 & 1.9    \\ 
AGN3 &143310.33+334604.5  &$<$3  &$<$0.02   &\nodata  &\nodata& emit   \\ 
AGN4 & 143409.54+334649.4   & $<$2 & $<$0.01  &10  &0.10$\pm$0.02 &emit   \\ 
AGN5 & 143105.67+341232.7 &\nodata   &\nodata    &\nodata   &\nodata  &emit?    \\ 
AGN6  & 143132.17+341417.9 &$<$5   &$<$0.03   &$<$3  &$<$0.02 &emit    \\ 
AGN7 & 143157.96+341650.1 & $<$2 &$<$0.01   &11  &0.06$\pm$0.01 &emit     \\

AGN8 &   143728.80+344547.6 & 9 &0.02$\pm$0.01   &$<$13  &$<$0.07 &0.51   \\ 
AGN9 & 143734.01+345721.3 &\nodata   & \nodata   &\nodata   &\nodata  &emit?   \\ 

AGN10 &142614.87+350616.5  &$<$6  &$<$0.06  &1.9  &0.02$\pm$0.01 &  emit?   \\ 

AGN11 &142827.20+354127.7 & $<$2 &$<$0.01   &\nodata  &\nodata &0.51   \\ 

AGN12 &171352.41+584201.2 &15  &0.04$\pm$0.01   &$<$9  &$<$0.05 & none?   \\ 

AGN13 &171207.43+584754.4  &$<$6  &$<$0.02   &4  &0.03$\pm$0.01 & 0.21   \\ 

AGN14 &171335.15+584756.1 &$<$5  &$<$0.01   &$<$6  &$<$0.03 &0.30   \\ 

AGN15 &171852.71+591432.0  &20 &0.37$\pm$0.05   &$<$5  &$<$0.03 &2.1     \\ 

AGN16 &171021.76+591854.7  & 4 &0.02$\pm$0.01   &$<$5  &$<$0.05 &0.29    \\ 

AGN17 &172825.80+592653.3 &\nodata   &\nodata    &\nodata   &\nodata  &abs?   \\ 
 
AGN18 &171839.73+593359.6  &$<$11  &$<$0.05   & 11 &0.11$\pm$0.05 & emit  \\ 
 
AGN19 &171302.37+593611.0  &$<$2  &$<$0.01   &$<$2  &$<$0.02 &3.2    \\ 

AGN20 & 171902.29+593715.9 &$<$9  & $<$0.02  &$<$3  &$<$0.01 & emit   \\   
 
AGN21 &172704.67+593736.6  &$<$10  & $<$0.01  &$<$10  & $<$0.03 & emit   \\ 

AGN22 &171530.75+600216.4  &$<$2  & $<$0.03  & 8  &0.16$\pm$0.04  & 0.64   \\

AGN23 &172123.19+601214.6 &\nodata  & \nodata  &\nodata  &\nodata  &\nodata   \\ 
 
AGN24 &171313.96+603146.6 &$<$9  & $<$0.02  &$<$9  & $<$0.05   &  emit  \\ 

\enddata

\tablenotetext{a}{Coordinates identify sources from Table 1 for FLS sources or from Table 1 in \citet{hou07} for Bootes sources. Source 11 is also in \citet{des06}.}
\tablenotetext{b}{Total flux of 6.2\,\um PAH emission feature in units of 10$^{-22}$W cm$^{-2}$, fit with single gaussian on a linear continuum within rest wavelength range 5.5\,\um to 6.9\,\um; uncertainties deriving from noise in the feature and from flux calibration are typically $\pm$ 10\%. Equivalent widths (EW) are in the rest frame.}
\tablenotetext{c}{Total flux of [NeIII] emission feature in units of 10$^{-22}$W cm$^{-2}$, fit with single gaussian on a linear continuum within rest wavelength range 14.9\,\um to 16.2\,\um; uncertainties deriving from noise in the feature and from flux calibration are typically $\pm$ 10\%. Equivalent widths (EW) are in the rest frame.}
\tablenotetext{d}{Optical depth $\tau_{si}$ is measure of depth of 9.7\,\um absorption feature, defined as $\tau_{si}$ = ln[$f_{\nu}$(cont)/$f_{\nu}$(abs)], for $f_{\nu}$(abs) measured at the maximum depth of absorption and $f_{\nu}$(cont) the unabsorbed continuum extrapolated from either side of the absorption feature as in \citet{spo07}. Strength of silicate emission features is not measured because of uncertainty in defining underlying continuum. }

\end{deluxetable}

\clearpage
\pagestyle{empty}

\begin{deluxetable}{lcclcccccc} 
\rotate
\tablecolumns{10}
\tabletypesize{\footnotesize}

\tablewidth{0pc}
\tablecaption{Redshifts and Continuum Luminosities for AGN in FLS and Bootes} 

\tablehead{

 \colhead{Source} & \colhead{Coordinate\tablenotemark{a}} & \colhead{z(IRS)} & \colhead {z(SDSS)\tablenotemark{b}}  & \colhead{f$_{\nu}$(5.5\ums)\tablenotemark{c}}& \colhead{f$_{\nu}$(8\ums)\tablenotemark{d}}& \colhead{f$_{\nu}$(15\ums)\tablenotemark{e}} & \colhead{log $\nu$L$_{\nu}$(15 \ums)\tablenotemark{f}} & \colhead{f$_{\nu}$(15\ums)/f$_{\nu}$(8\ums)} & \colhead{f$_{\nu}$(24\ums)/f$_{\nu}$(8\ums)}\\ 

\colhead{} & \colhead {} & \colhead{} & \colhead{} &\colhead {mJy} &\colhead {mJy} &\colhead{} &\colhead{} &\colhead{} &\colhead{}
}

\startdata



AGN1 &   143156.40+325138.1 & \nodata & 0.4120(Sy2) &1.7 &3.1 & 10.5 &44.95 & 3.4 & 5.4  \\ 
AGN2 &  143205.63+325835.2 &0.48  &\nodata  &1.5 & 5.7 &15.9  &45.27  &2.8 &8.8  \\ 
AGN3 &  143310.33+334604.5  & \nodata & 2.4(QSO1)\tablenotemark{g} &9.1  &12.7  &\nodata &\nodata &\nodata &\nodata \\
AGN4 &   143409.54+334649.4 &0.216  & \nodata & 1.5 & 3.1 &8.9  &44.29  &  2.9  & 5.1 \\
AGN5  &  143105.67+341232.7 &0.5?  &\nodata  &\nodata   & \nodata & \nodata  & \nodata & \nodata \\ 
AGN6 &   143132.17+341417.9  &\nodata  &1.037(QSO1)  &4.8  & 7.0 &16.1  & 45.96 & 2.3 & \nodata  \\  
AGN7  &   143157.96+341650.1 &0.72  &0.7155(QSO1)  &4.4  &7.8 &23.3  &45.79  & 3.0  &\nodata  \\ 
AGN8 &   143728.80+344547.6  &0.57  &\nodata  &6.6 &14.1  &24.1  &45.60  &1.7  &\nodata    \\ 
AGN9 &   143734.01+345721.3  &0.5?  &\nodata  &\nodata  &\nodata &\nodata   &\nodata   &\nodata   &\nodata   \\
AGN10 &  142614.87+350616.5  &\nodata  &0.2168(Sy1)  & 1.4 &3.0 & 8.1  &44.25  &2.7 &5.3 \\ 
AGN11 &    142827.20+354127.7 & 1.30 & 1.293 & 9.4  & 13.0 & 19.2 &46.22   & 2.0   & \nodata   \\

AGN12 &171352.41+584201.2  &0.51  &0.5209(QSO)  &5.5 &9.6 &22.2  &45.49  &2.31  &\nodata  \\ 

AGN13 &171207.43+584754.4  &\nodata  & 0.2693(NLSy1) & 3.5 & 5.7 &10.9  &44.58  &1.9   &2.4   \\ 
AGN14 &171335.15+584756.1  &0.12  &0.1340(Sy1)  &5.6  &9.1 &15.5  &44.10  &1.7   & 3.0  \\ 

AGN15 &171852.71+591432.0   &0.329  &0.3220(liner)  & 5.3 &12.7 &0.24  &44.83  &2.4 &6.1   \\  

AGN16 &171021.76+591854.7   &0.417  &0.4186(Sy1)  &4.3 &7.2 &12.2  &45.02  & 1.7  & 3.8  \\ 

AGN17 &172825.80+592653.3  &0.3?  &\nodata  & \nodata   & \nodata   & \nodata   & \nodata   & \nodata   & \nodata   \\  
 
AGN18 &171839.73+593359.6  &\nodata  &0.3825(QSO)  &2.6  &4.1 &11.1  &44.90 & 2.7 &4.3  \\ 
 
AGN19 &171302.37+593611.0  &0.684  &0.668(AGN)  &3.8 & 7.4 &11.1  &45.43  & 1.5  & \nodata  \\ 

AGN20 & 171902.29+593715.9 &\nodata  &0.1783(QSO)  &5.7 & 9.1 &20.0  &44.47  & 2.2  &3.2  \\ 
 
AGN21 &172704.67+593736.6  &\nodata  &1.1284(QSO)  &10.0 &14.7  &27.9  &46.27  &1.9   &\nodata  \\ 

AGN22 &171530.75+600216.4  &0.43  &0.4204(Sy1)  &0.70 & 2.2 &9.1  &44.90  & 4.1  & 7.1  \\ 

AGN23 &172123.19+601214.6 &1.09?  &0.35? &\nodata  & \nodata &\nodata  &\nodata  &\nodata  &\nodata  \\  
AGN24 &171313.96+603146.6 &\nodata  &0.1050(AGN)  &3.4 & 5.1 &7.8  &43.59  &1.5  & 2.0  \\

\enddata
\tablenotetext{a}{Coordinates identify sources from Table 1 for FLS sources or from Table 1 in \citet{hou07} for Bootes sources.}
\tablenotetext{b}{Optical redshift from SDSS and classification, in parenthesis, of optical spectrum as given in NED archive.} 
\tablenotetext{c}{Continuum flux density at 5.5\,\um, measured as median flux between 5.0\,\um and 6.0\,\um; uncertainties are typically $\pm$ 5\%, based on uncertainties in flux calibration.}
\tablenotetext{d}{Continuum flux density at 8\,\um used for normalizing spectra, as described in text.}
\tablenotetext{e}{Continuum flux density at 15\,\um, measured as median flux between 14.8\,\um and 15.2\,\um; uncertainties are typically $\pm$ 5\%, based on uncertainties in flux calibration.}
\tablenotetext{f}{Rest frame luminosity $\nu$L$_{\nu}$(15$\mu$m) in ergs s$^{-1}$ determined using f$_{\nu}$(15$\mu$m) and luminosity distances from E.L. Wright, http://www.astro.ucla.edu/~wright/CosmoCalc.html, for H$_0$ = 71 \kmsMpc, $\Omega_{M}$=0.27 and $\Omega_{\Lambda}$=0.73.  (Log [$\nu$L$_{\nu}$(15$\mu$m)(\ldot)] = log [$\nu$L$_{\nu}$(15$\mu$m)(ergs s$^{-1}$)] - 33.59.)} 
\tablenotetext{g}{Redshift from \citet {bro06} using AGES survey.}

\end{deluxetable}

\clearpage
\pagestyle{empty}

\begin{deluxetable}{ccccc} 

\tablecolumns{5}
\tabletypesize{\footnotesize}

\tablewidth{0pc}
\tablecaption{Average Spectra for Different Luminosities} 

\tablehead{

 \colhead{wavelength} & \colhead{42.0$<$log$L$$<$43.0\tablenotemark{a}} &\colhead{43.0$<$log$L$$<$44.0\tablenotemark{a} }& \colhead{44.0$<$log$L$$<$45.0\tablenotemark{a} }& \colhead{log$L$$>$45.0\tablenotemark{a}}\\ 

\colhead{\ums} & \colhead {normalized L$_{\nu}$} & \colhead{normalized L$_{\nu}$} & \colhead{normalized L$_{\nu}$} &\colhead {normalized L$_{\nu}$}
}

\startdata
5.0 & 0.06 & 0.07 & 0.30 & 0.49 \\
5.2 & 0.07 & 0.08 & 0.32 & 0.51 \\
5.4 & 0.08 & 0.10 & 0.35 & 0.53 \\
5.6 & 0.11 & 0.15 & 0.38 & 0.55 \\
5.8 & 0.13 & 0.18 & 0.41 & 0.58 \\
6.0 & 0.18 & 0.22 & 0.43 & 0.60 \\
6.2 & 0.49 & 0.54 & 0.60 & 0.63 \\
6.4 & 0.34 & 0.33 & 0.57 & 0.66 \\
6.6 & 0.27 & 0.30 & 0.54 & 0.68 \\
6.8 & 0.24 & 0.32 & 0.58 & 0.71 \\
7.0 & 0.30 & 0.40 & 0.64 & 0.72 \\
7.2 & 0.40 & 0.42 & 0.66 & 0.76 \\
7.4 & 0.49 & 0.62 & 0.79 & 0.85 \\
7.6 & 0.77 & 0.99 & 0.98 & 0.95 \\
7.7 & 1.00 & 1.00 & 1.00 & 1.00 \\
7.8 & 0.96 & 0.95 & 1.00 & 1.00 \\
8.0 & 0.64 & 0.75 & 0.91 & 0.98 \\
8.2 & 0.52 & 0.63 & 0.84 & 0.96 \\
8.4 & 0.54 & 0.66 & 0.83 & 0.93 \\
8.6 & 0.62 & 0.69 & 0.83 & 0.91 \\
8.8 & 0.47 & 0.51 & 0.78 & 0.89 \\
9.0 & 0.36 & 0.35 & 0.75 & 0.89 \\
9.2 & 0.34 & 0.33 & 0.71 & 0.88 \\
9.4 & 0.29 & 0.32 & 0.74 & 0.91 \\
9.6 & 0.22 & 0.32 & 0.79 & 0.94 \\
9.8 & 0.24 & 0.32 & 0.81 & 0.94 \\
10.0 & 0.24 & 0.32 & 0.83 & 0.95 \\
10.2 & 0.21 & 0.33 & 0.85 & 0.99 \\
10.4 & 0.36 & 0.37 & 0.92 & 1.02 \\
10.6 & 0.33 & 0.43 & 1.00 & 1.05 \\
10.8 & 0.36 & 0.42 & 1.06 & 1.07 \\
11.0 & 0.49 & 0.62 & 1.11 & 1.10 \\
11.2 & 1.04 & 1.14 & 1.43 & 1.17 \\
11.3 & 1.27 & 1.22 & 1.46 & 1.24 \\
11.4 & 1.03 & 1.01 & 1.42 & 1.27 \\
11.6 & 0.72 & 0.68 & 1.35 & 1.24 \\
11.8 & 0.59 & 0.87 & 1.37 & 1.29 \\
12.0 & 0.63 & 0.77 & 1.41 & 1.34 \\
12.2 & 0.72 & 0.81 & 1.55 & 1.39 \\
12.4 & 0.72 & 0.91 & 1.63 & 1.49 \\
12.6 & 0.88 & 1.05 & 1.73 & 1.55 \\
12.7 & 0.90 & 1.21 & 1.80 & 1.59 \\
12.8 & 0.83 & 1.22 & 1.82 & 1.62 \\
13.0 & 0.71 & 0.98 & 1.75 & 1.66 \\
13.2 & 0.66 & 0.91 & 1.71 & 1.70 \\
13.4 & 0.63 & 0.84 & 1.63 & 1.73 \\
13.6 & 0.57 & 0.79 & 1.66 & 1.76 \\
13.8 & 0.56 & 0.80 & 1.71 & 1.80 \\
14.0 & 0.55 & 0.78 & 1.76 & 1.82 \\
14.2 & 0.65 & 0.82 & 1.84 & 1.93 \\
14.4 & 0.68 & 0.81 & 1.90 & 2.00 \\
14.6 & 0.69 & 0.80 & 1.95 & 2.01 \\
14.8 & 0.71 & 0.80 & 2.04 & 2.01 \\
15.0 & 0.71 & 0.82 & 2.04 & 2.04 \\
15.2 & 0.72 & 0.84 & 2.07 & 2.06 \\
15.4 & 0.72 & 0.91 & 2.22 & 2.16 \\
15.6 & 1.14 & 1.05 & 2.40 & 2.20 \\
15.7 & 1.14 & 0.96 & 2.46 & 2.21 \\
15.8 & 0.96 & 0.94 & 2.33 & 2.21 \\
16.0 & 0.87 & 1.00 & 2.29 & 2.22 \\
16.2 & 0.88 & 1.08 & 2.35 & 2.23 \\
16.4 & 0.94 & 1.21 & 2.42 & 2.25 \\
16.6 & 1.10 & 1.24 & 2.47 & 2.30 \\
16.8 & 1.28 & 1.33 & 2.55 & 2.34 \\
17.0 & 1.41 & 1.41 & 2.69 & 2.38 \\
17.2 & 1.44 & 1.40 & 2.79 & 2.43 \\
17.4 & 1.50 & 1.38 & 2.83 & 2.46 \\
17.6 & 1.36 & 1.32 & 2.86 & 2.49 \\
17.8 & 1.27 & 1.34 & 2.89 & 2.56 \\
18.0 & 1.21 & 1.38 & 2.90 & 2.60 \\
18.2 & 1.25 & 1.44 & 2.90 & 2.63 \\
18.4 & 1.29 & 1.47 & 2.91 & 2.67 \\
18.6 & 1.32 & 1.52 & 2.96 & 2.78 \\
18.8 & 1.38 & 1.56 & 3.05 & 2.79 \\
19.0 & 1.54 & 1.55 & 3.16 & 2.82 \\
19.2 & 1.62 & 1.54 & 3.14 & 2.89 \\
19.4 & 1.71 & 1.56 & 3.13 & 2.96 \\
19.6 & 1.70 & 1.60 & 3.19 & 3.04 \\
19.8 & 1.68 & 1.64 & 3.25 & 3.10 \\
20.0 & 1.70 & 1.73 & 3.29 & 3.19 \\
20.2 & 1.80 & 1.74 & 3.44 & 3.30 \\
20.4 & 1.90 & 1.75 & 3.47 & 3.44 \\
20.6 & 1.96 & 1.77 & 3.48 & 3.56 \\
20.8 & 2.00 & 1.85 & 3.47 & 3.67 \\
21.0 & 2.04 & 1.91 & 3.50 & 3.70 \\
21.2 & 2.06 & 1.95 & 3.52 & 3.75 \\
21.4 & 2.09 & 2.00 & 3.52 & 3.92 \\
21.6 & 2.11 & 2.01 & 3.59 & 4.11 \\
21.8 & 2.15 & 2.03 & 3.70 & 4.22 \\
22.0 & 2.19 & 2.07 & 3.77 & 4.34 \\
22.2 & 2.21 & 2.12 & 3.85 & 4.50 \\
22.4 & 2.25 & 2.17 & 3.88 & 4.61 \\
22.6 & 2.32 & 2.23 & 3.91 & 4.72 \\
22.8 & 2.48 & 2.32 & 3.99 & 4.87 \\
23.0 & 2.54 & 2.39 & 4.05 & 5.28 \\
23.2 & 2.56 & 2.42 & 4.10 & 5.50 \\
23.4 & 2.57 & 2.48 & 4.14 & 5.75 \\
23.6 & 2.58 & 2.53 & 4.20 & 5.99 \\
23.8 & 2.62 & 2.58 & 4.24 & 6.02 \\
24.0 & 2.67 & 2.64 & 4.28 & 6.05 \\
24.2 & 2.75 & 2.66 & 4.33 & 6.10 \\
24.4 & 2.78 & 2.73 & 4.39 & 6.15 \\
24.6 & 2.80 & 2.77 & 4.42 & 6.22 \\
24.8 & 2.83 & 2.81 & 4.45 & 6.26 \\
25.0 & 2.85 & 2.87 & 4.51 & 6.32 \\

\enddata
\tablenotetext{a}{Luminosity $L$ is defined as $L$ = $\nu$L$_{\nu}$(15$\mu$m) in ergs s$^{-1}$ in the source rest frame. (Log [$\nu$L$_{\nu}$(15$\mu$m)(\ldot)] = log [$\nu$L$_{\nu}$(15$\mu$m)(ergs s$^{-1}$)] - 33.59.)}
\end{deluxetable}
\clearpage
%
%

\begin{figure}
\figurenum{1}
 
\includegraphics[scale=0.9]{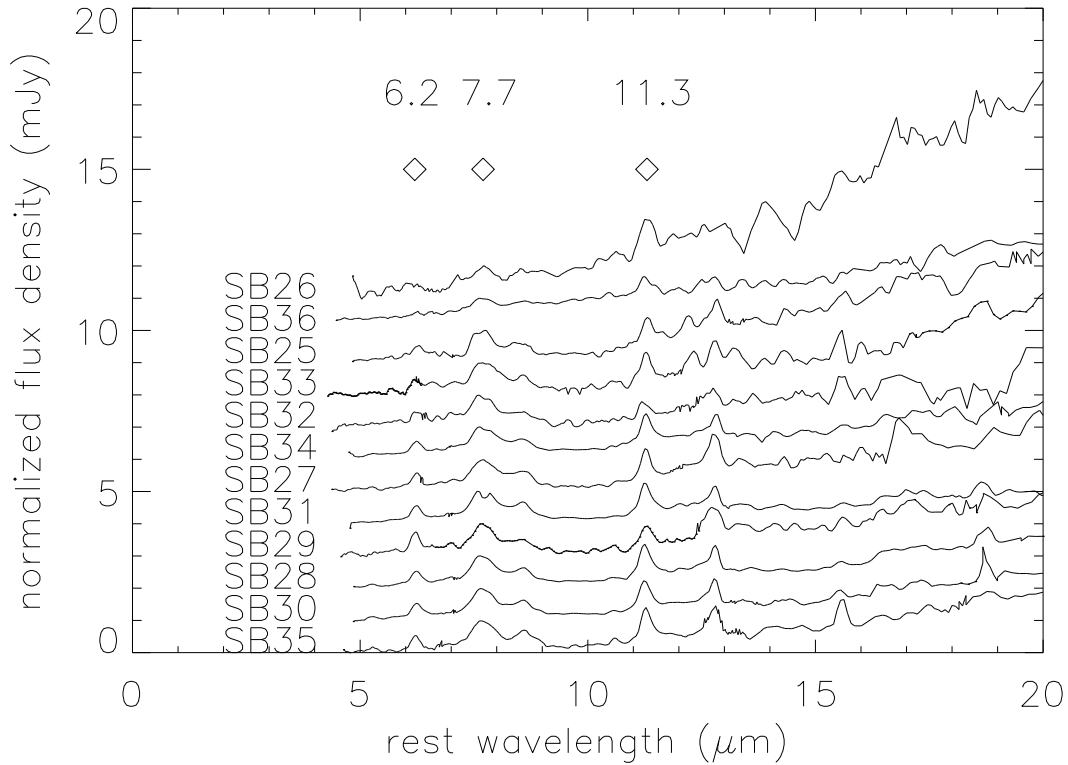}
\caption{Rest-frame spectra of all FLS sources with PAH features in Tables 2 and 3, truncated at 20\,\um because there are no spectral features beyond that wavelength. All spectra are normalized to peak f$_{\nu}$(7.7\ums) = 1.0 mJy, but zero points are displaced for illustration.  The zero flux level for each spectrum is located 1 mJy below the peak at 7.7$\mu$m. Results show the similarity among all starburst spectra; diamonds show positions of strong PAH features at 6.2\ums, 7.7\ums, and 11.3\ums.  Spectra are labeled by running numbers for starbursts in Tables 2 and 3.}

\end{figure}

\begin{figure}
\figurenum{2}
 
\includegraphics[scale=0.9]{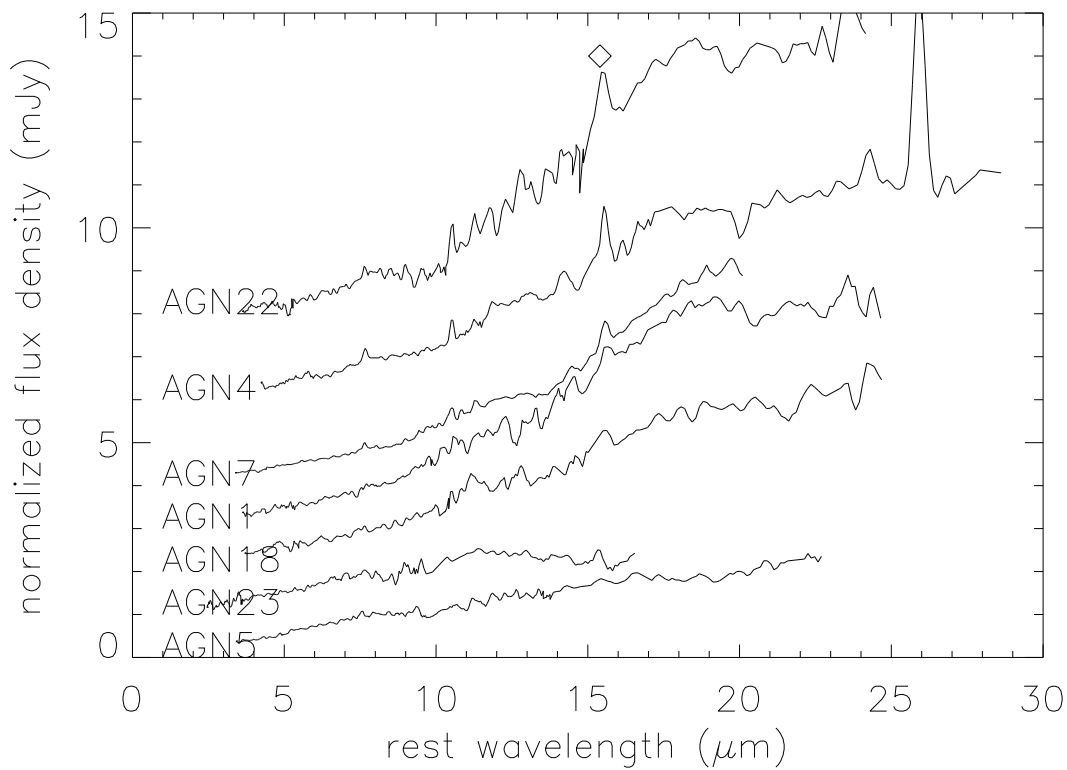}
\caption{Rest-frame spectra of AGN sources with silicate emission features in Tables 4 and 5 which also have a measurable [NeIII] 15.6$\mu$m emission line (position shown by diamond). All spectra are normalized to f$_{\nu}$(8\ums) = 1.0 mJy, but zero points are displaced for illustration.  The zero flux level for each spectrum is located 1 mJy below the value at 8$\mu$m. Spectra are labeled by running numbers for AGN in Tables 4 and 5. Source AGN5 has uncertain redshift and estimated redshift of 0.5 which is shown and given in Table 5 derives from presumed weak 15.6$\mu$m feature. Source AGN23 is highly uncertain, being shown with IRS redshift of 1.09 derived from assumed 15.6$\mu$m feature and similarity of silicate emission to source AGN13 in Figure 3, but this redshift disagrees with the SDSS redshift of 0.35.}

\end{figure}

\begin{figure}
\figurenum{3}
 
\includegraphics[scale=0.9]{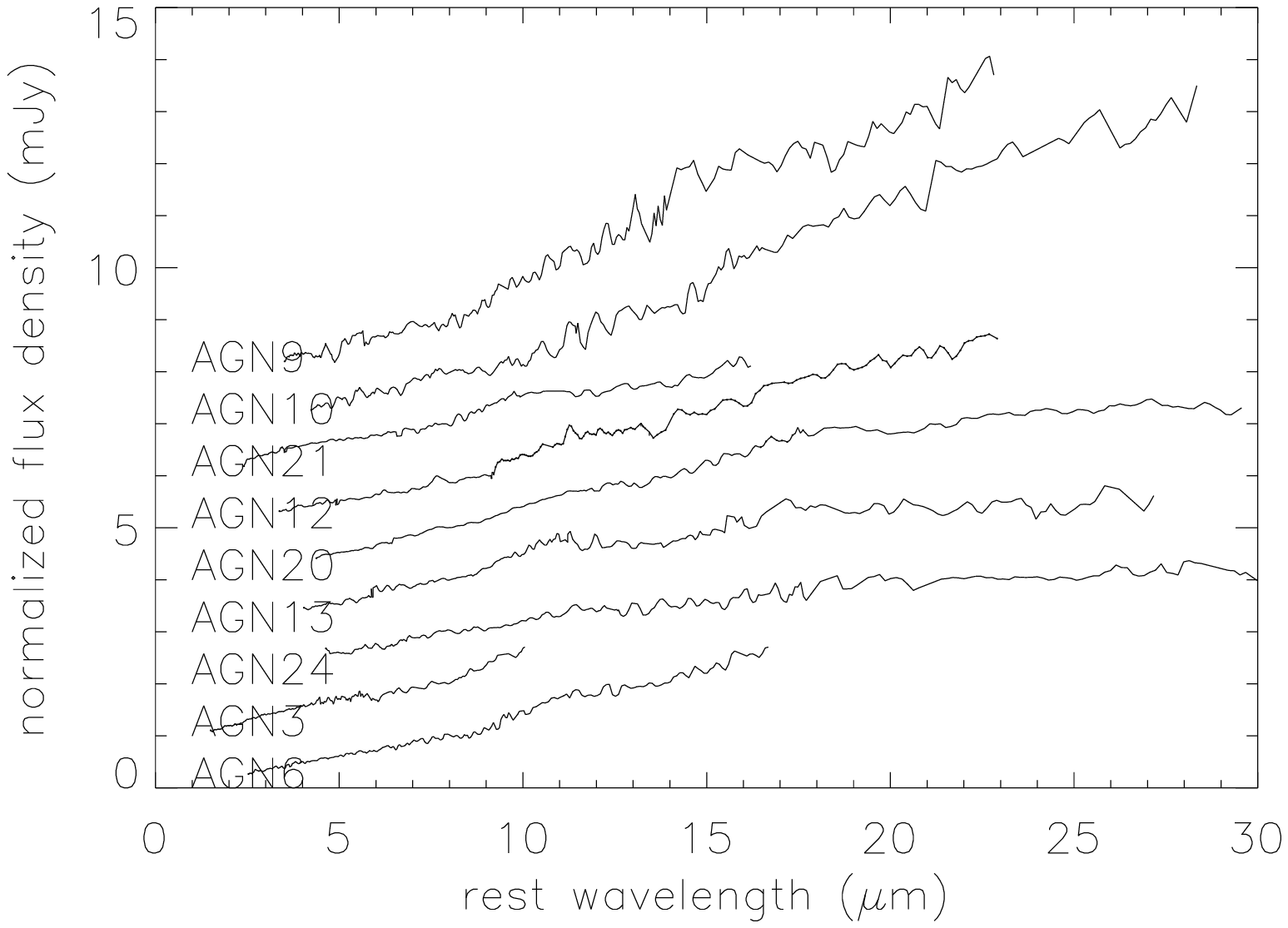}
\caption{Rest-frame spectra of remaining AGN sources with silicate emission features in Tables 4 and 5 not shown in Figure 2.  All spectra are normalized to f$_{\nu}$(8\ums) = 1.0 mJy, but zero points are displaced for illustration.  The zero flux level for each spectrum is located 1 mJy below the value at 8$\mu$m. Spectra are labeled by running numbers for AGN in Tables 4 and 5. Source AGN9 has uncertain redshift; estimated redshift of 0.5 which is shown and given in Table 5 derives from presumed silicate feature at 17\ums.}

\end{figure}

\begin{figure}
\figurenum{4}
 
\includegraphics[scale=0.9]{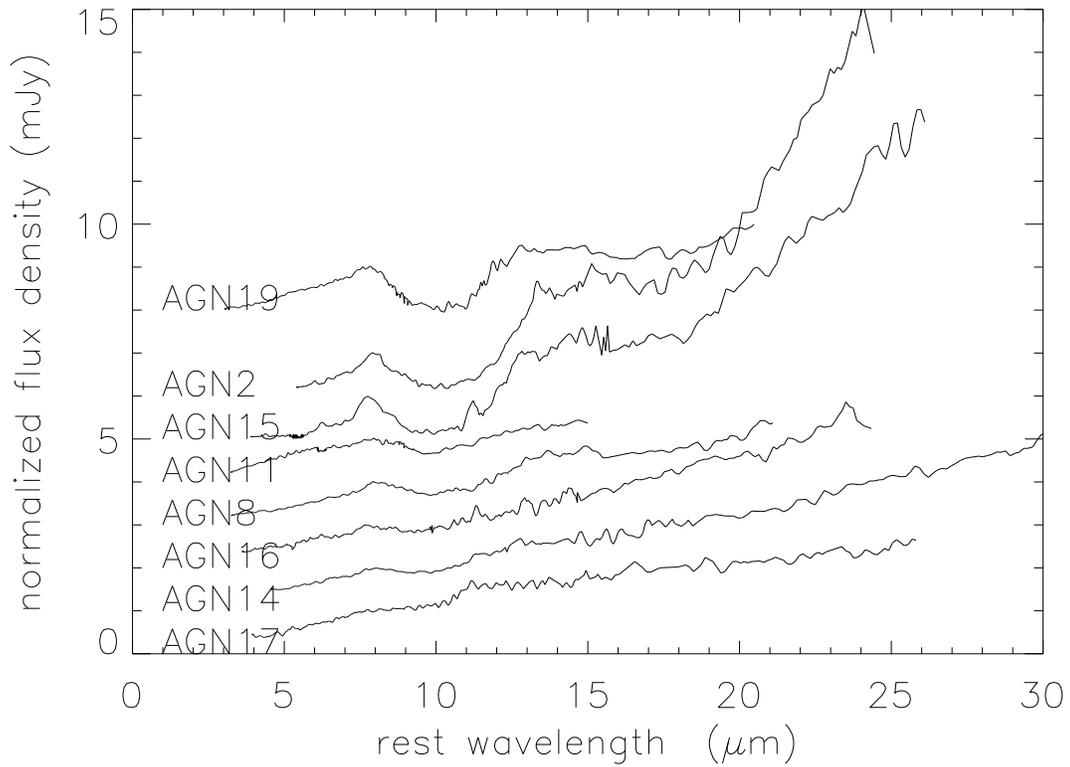}
\caption{Rest-frame spectra of all AGN sources with silicate absorption features in Tables 4 and 5. All spectra are normalized to f$_{\nu}$(8\ums) = 1.0 mJy, but zero points are displaced for illustration.  The zero flux level for each spectrum is located 1 mJy below the value at 8$\mu$m. Spectra are labeled by running numbers for AGN in Tables 4 and 5. Source AGN17 has uncertain redshift; estimated redshift of 0.3 which is shown and given in Table 5 derives from presumed silicate 9.7\,\um absorption feature.}
\end{figure}

\begin{figure}
\figurenum{5}
\includegraphics[scale=0.9]{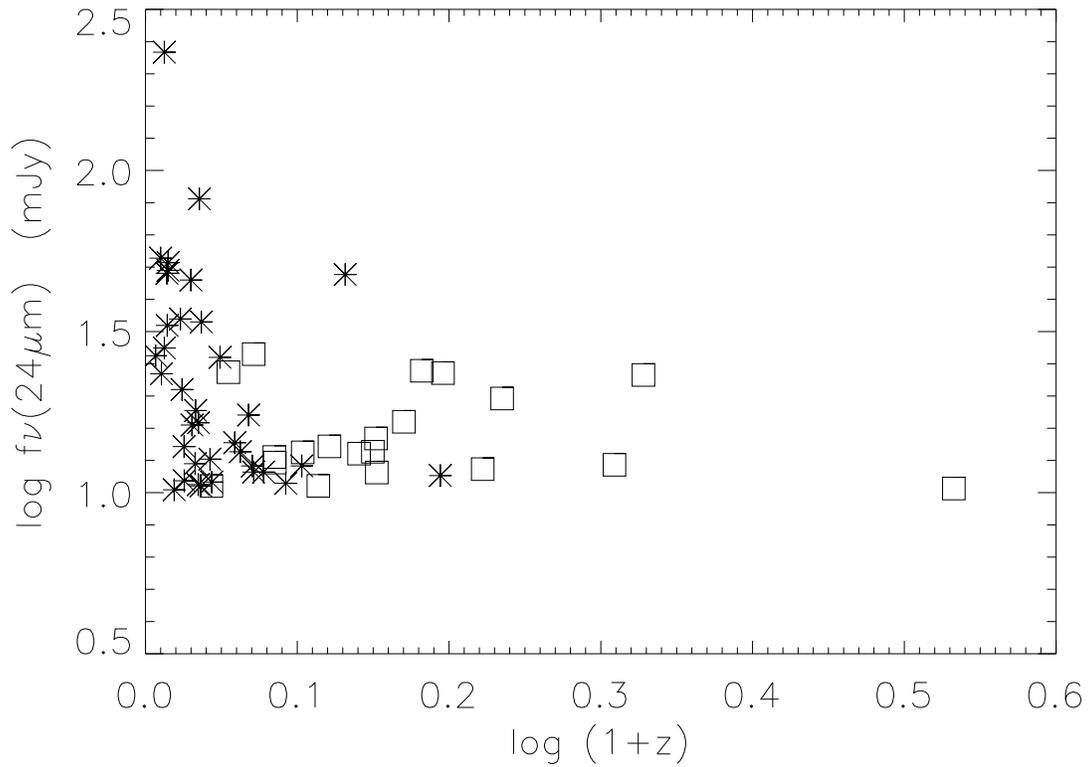}
\caption{Comparison of f$_{\nu}$(24\ums) with redshift for all Bootes and FLS sources.  Asterisks indicate starbursts in Tables 2 and 3; squares indicate AGN in Tables 4 and 5.  Typically higher luminosities of AGN sources allow their discovery to higher redshifts; the source of highest redshift has z = 2.4.} 
\end{figure}

\begin{figure}
\figurenum{6}
\includegraphics[scale=0.9]{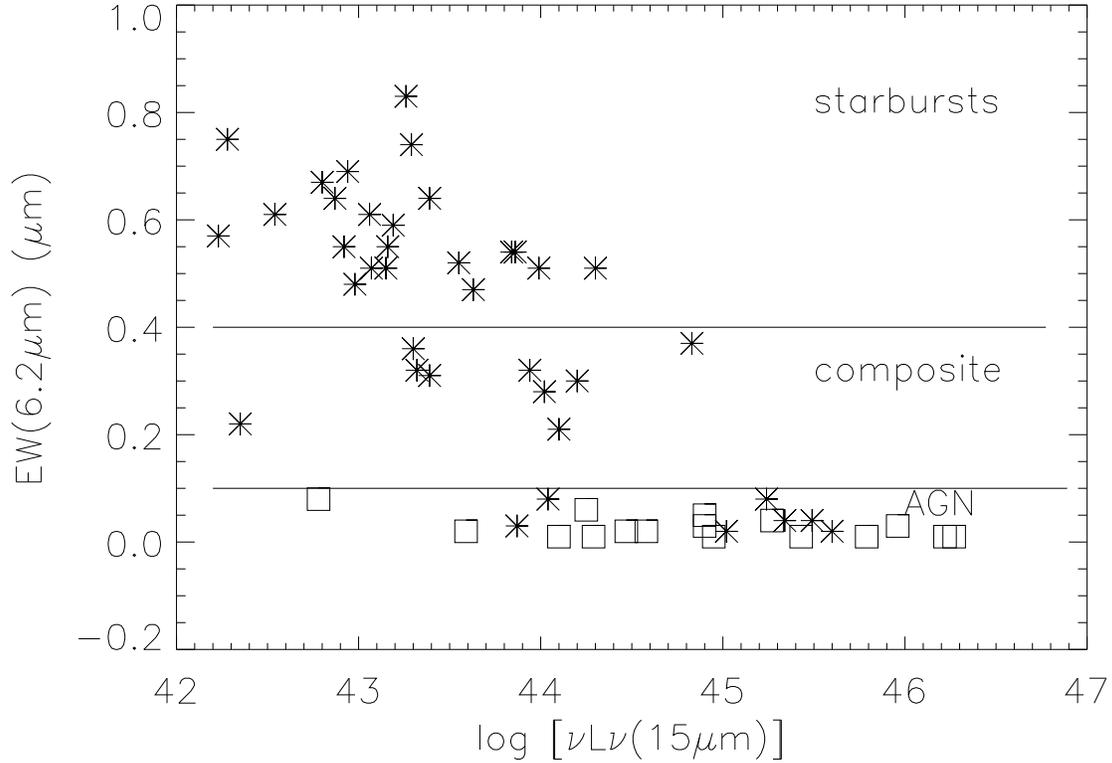}
\caption{Comparison of continuum luminosity $\nu$L$_{\nu}$(15$\mu$m) in ergs s$^{-1}$ with rest-frame equivalent width of 6.2\,\um PAH feature for all Bootes and FLS sources.  (Log [$\nu$L$_{\nu}$(15$\mu$m)(\ldot)] = log [$\nu$L$_{\nu}$(15$\mu$m)(ergs s$^{-1}$)] - 33.59.)  Asterisks indicate sources with measurable PAH emission; squares show EW limits for sources without measurable PAH, considered to be pure AGN. Results show the larger luminosities arising from the AGN sources.  The break in the distribution at EW(6.2\ums) = 0.4\um is interpreted as dividing pure starbursts without any AGN contribution, having EW(6.2\ums) $>$ 0.4\ums, from sources with starburst plus AGN composite spectra. } 
\end{figure}

\begin{figure}
\figurenum{7}
 
\includegraphics[scale=0.9]{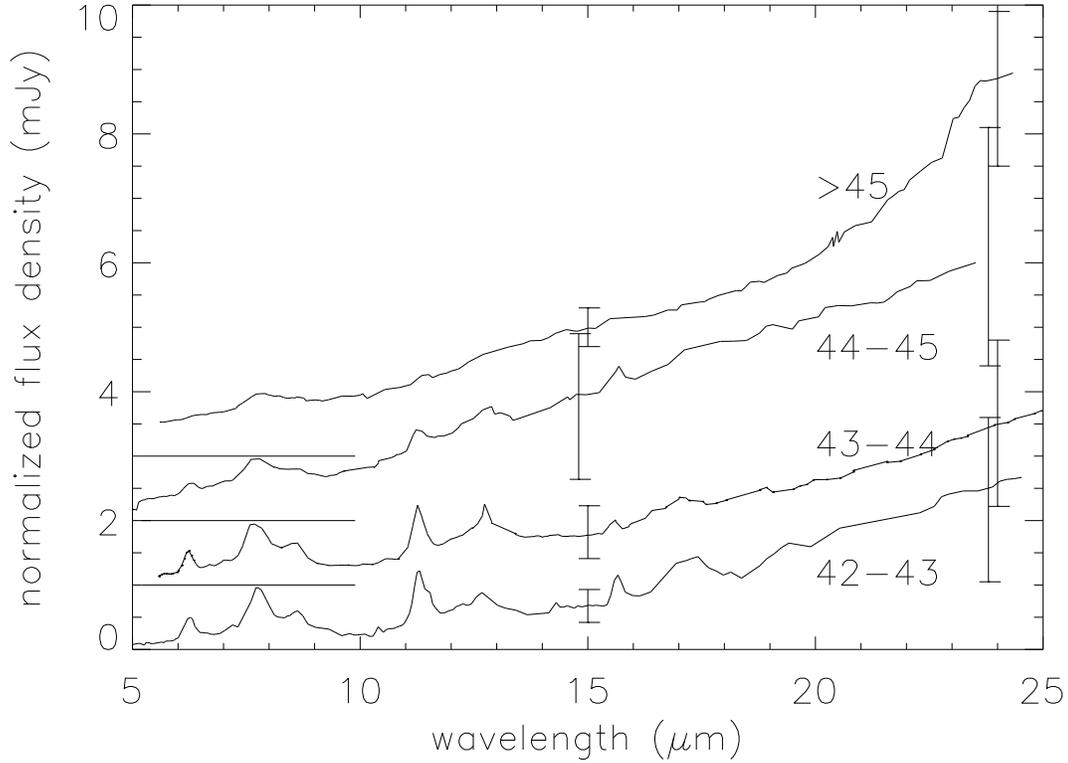}
\caption{Average spectra of sources within different luminosity intervals of log $\nu$L$_{\nu}$(15$\mu$m).  All spectra are normalized to 1 mJy flux density at 7.7\,\um and displaced for illustration.  Zero flux density for each spectrum is shown by short horizontal lines.  From top down, average spectra are for log $\nu$L$_{\nu}$(15$\mu$m) $>$ 45.0 (ergs s$^{-1}$), 44.0 $<$ log $\nu$L$_{\nu}$(15$\mu$m) $<$ 45.0, 43.0 $<$ log $\nu$L$_{\nu}$(15$\mu$m) $<$ 44.0, and 42.0 $<$ log $\nu$L$_{\nu}$(15$\mu$m) $<$ 43.0.  (Log [$\nu$L$_{\nu}$(15$\mu$m)(\ldot)] = log [$\nu$L$_{\nu}$(15$\mu$m)(ergs s$^{-1}$)] - 33.59.)  Error bars indicate 1 $\sigma$ dispersion in continuum slopes for sources that enter average, as displayed in Figures 8-11.  Average spectra for fainter luminosities are dominated by PAH features from starbursts. }
\end{figure}

\begin{figure}
\figurenum{8}
 
\includegraphics[scale=0.9]{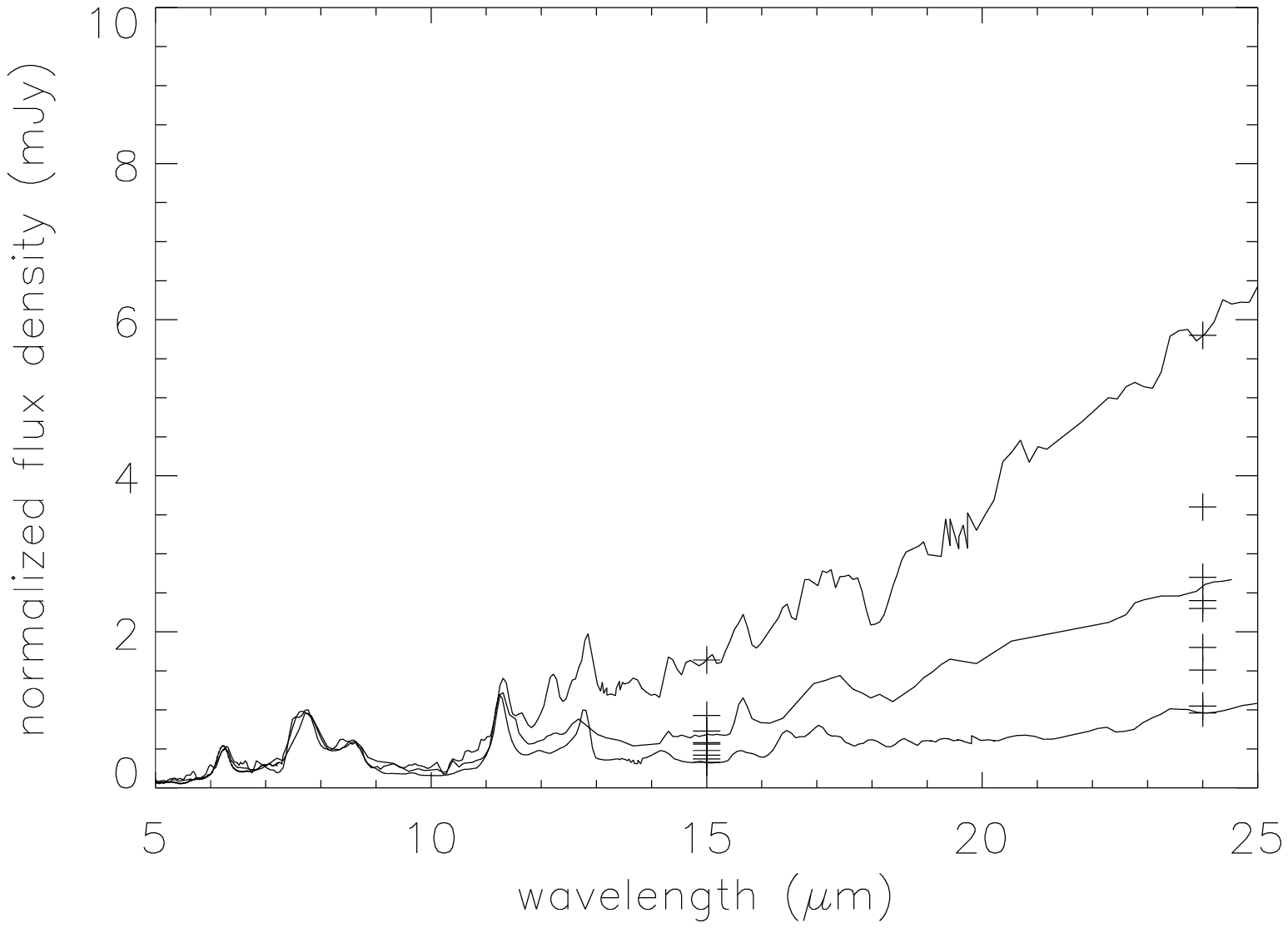}
\caption{Average spectrum of sources with 42.0 $<$ log $\nu$L$_{\nu}$(15$\mu$m) $<$ 43.0 (center spectrum) normalized to f$_{\nu}$(7.7\ums) = 1.0 mJy; extreme spectra within this luminosity bin are shown as measured at 24$\mu$m; upper curve is source SB25 in Tables 2 and 3, and lower curve is source SB23 in Tables 2 and 3.  Crosses show dispersions in spectral shape for all individual sources that enter the average, showing continuum fluxes at 15$\mu$m and 24$\mu$m normalized to f$_{\nu}$(7.7\ums) = 1.0 mJy.} 
\end{figure}

\begin{figure}
\figurenum{9}
 
\includegraphics[scale=0.9]{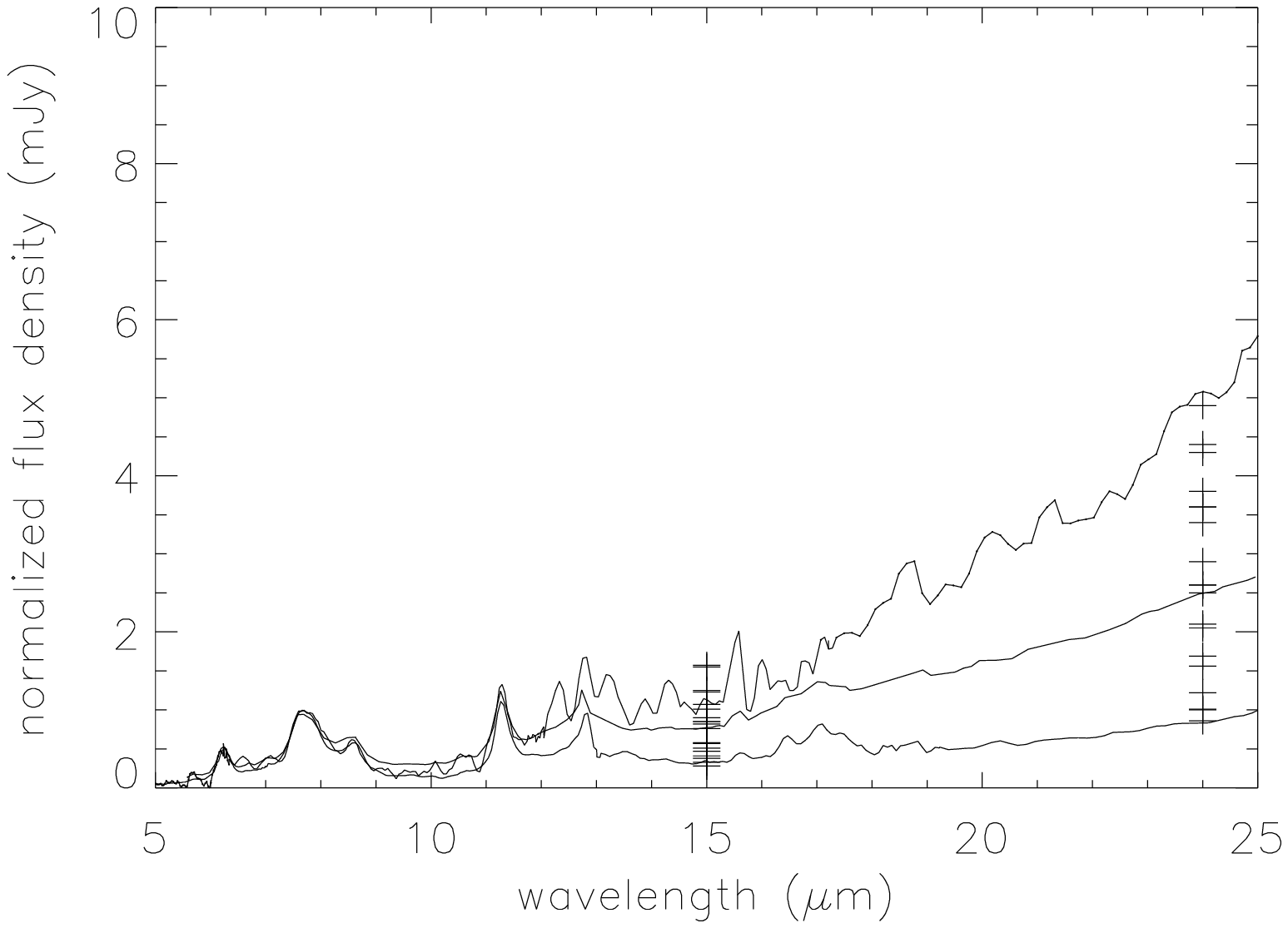}
\caption{Average spectrum of sources with 43.0 $<$ log $\nu$L$_{\nu}$(15$\mu$m) $<$ 44.0 (center spectrum) normalized to f$_{\nu}$(7.7\ums) = 1.0 mJy; extreme spectra within this luminosity bin are shown as measured at 24$\mu$m; upper curve is source SB33 in Tables 2 and 3, and lower curve is source SB16 in Tables 2 and 3.  Crosses show dispersions in spectral shape for individual sources that enter the average, showing continuum fluxes at 15$\mu$m and 24$\mu$m normalized to f$_{\nu}$(7.7\ums) = 1.0 mJy.} 
\end{figure}

\begin{figure}
\figurenum{10}
 
\includegraphics[scale=0.9]{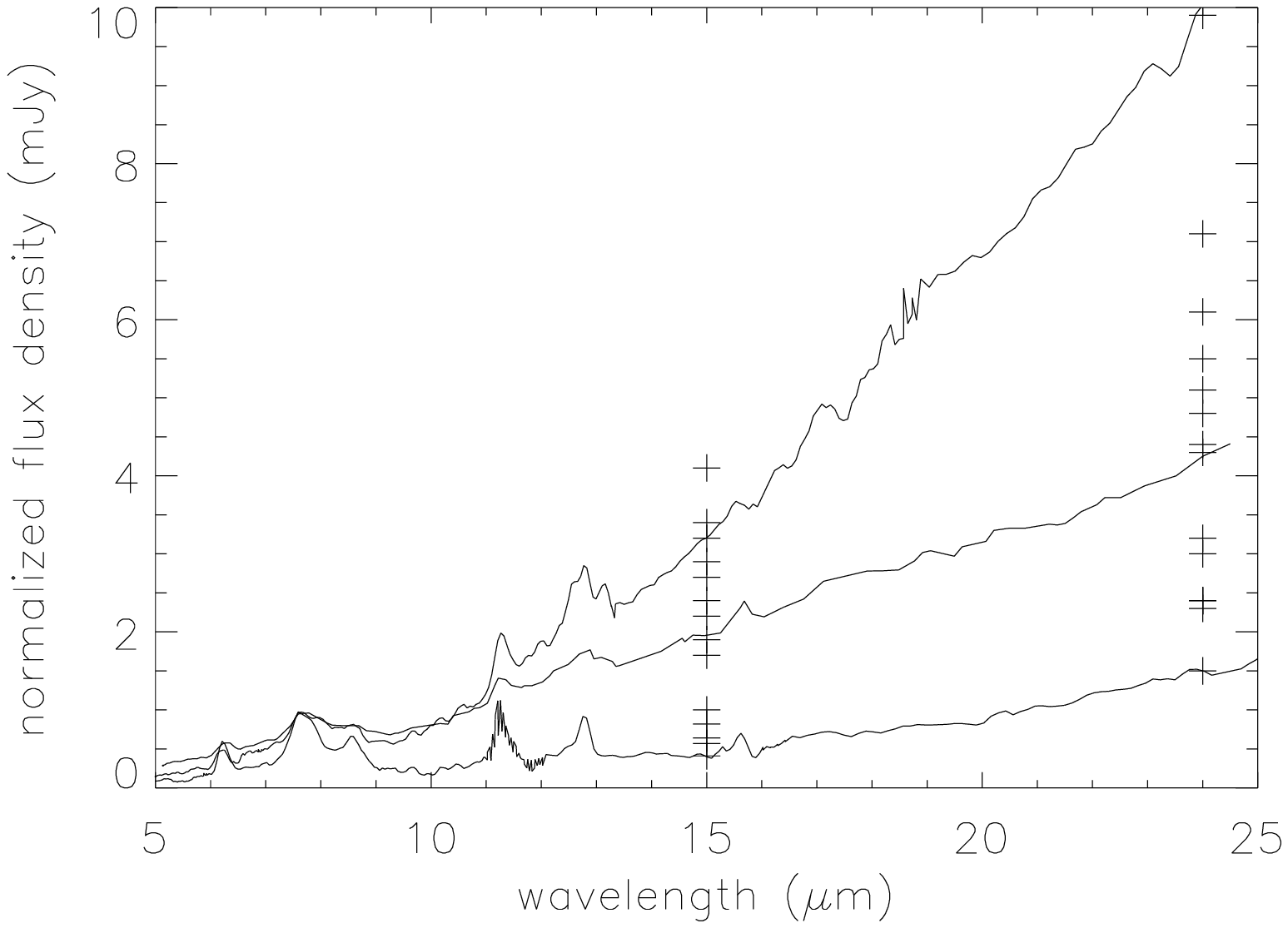}
\caption{Average spectrum of sources with 44.0 $<$ log $\nu$L$_{\nu}$(15$\mu$m) $<$ 45.0 (center spectrum) normalized to f$_{\nu}$(7.7\ums) = 1.0 mJy; extreme spectra within this luminosity bin are shown as measured at 24$\mu$m; upper curve is source SB18 in Tables 2 and 3, and lower curve is source SB1 in Tables 2 and 3.  Crosses show dispersions in spectral shape for individual sources that enter the average, showing continuum fluxes at 15$\mu$m and 24$\mu$m normalized to f$_{\nu}$(7.7\ums) = 1.0 mJy. This luminosity bin shows the greatest dispersion about the average of all luminosity bins showns in Figures 8-11, primarily because this bin contains the most varied assortment of sources, including luminous starbursts and heavily absorbed AGN.}
\end{figure}

\begin{figure}
\figurenum{11}
 
\includegraphics[scale=0.9]{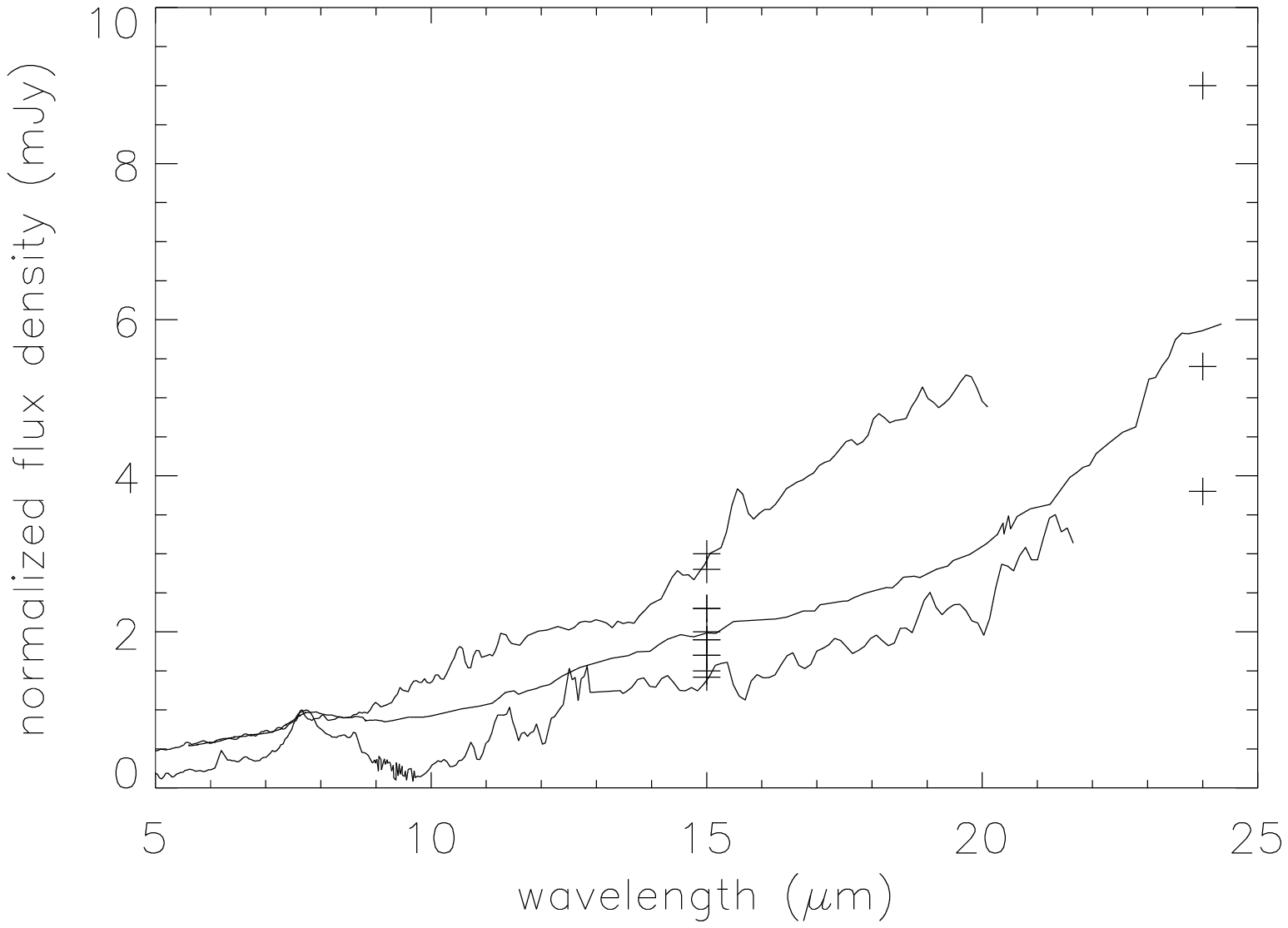}
\caption{Average spectrum of sources with log $\nu$L$_{\nu}$(15$\mu$m) $>$ 45.0 (center spectrum) normalized to f$_{\nu}$(7.7\ums) = 1.0 mJy; extreme spectra within this luminosity bin are shown as measured at 15$\mu$m; upper curve is source AGN7 in Tables 4 and 5, and lower curve is source SB10 in Tables 2 and 3.  Crosses show dispersions in spectral shape for individual sources that enter the average, showing continuum fluxes at 15$\mu$m and 24$\mu$m normalized to f$_{\nu}$(7.7\ums) = 1.0 mJy. Note that average for $\lambda$ $>$ 20$\mu$m is uncertain because only 3 sources have sufficiently low redshifts for rest frame spectra to be seen at these wavelengths; extremes for this Figure are determined at 15$\mu$m because so few sources have rest-frame measures at 24$\mu$m.}
\end{figure}

\begin{figure}
\figurenum{12}
 
\includegraphics[scale=0.9]{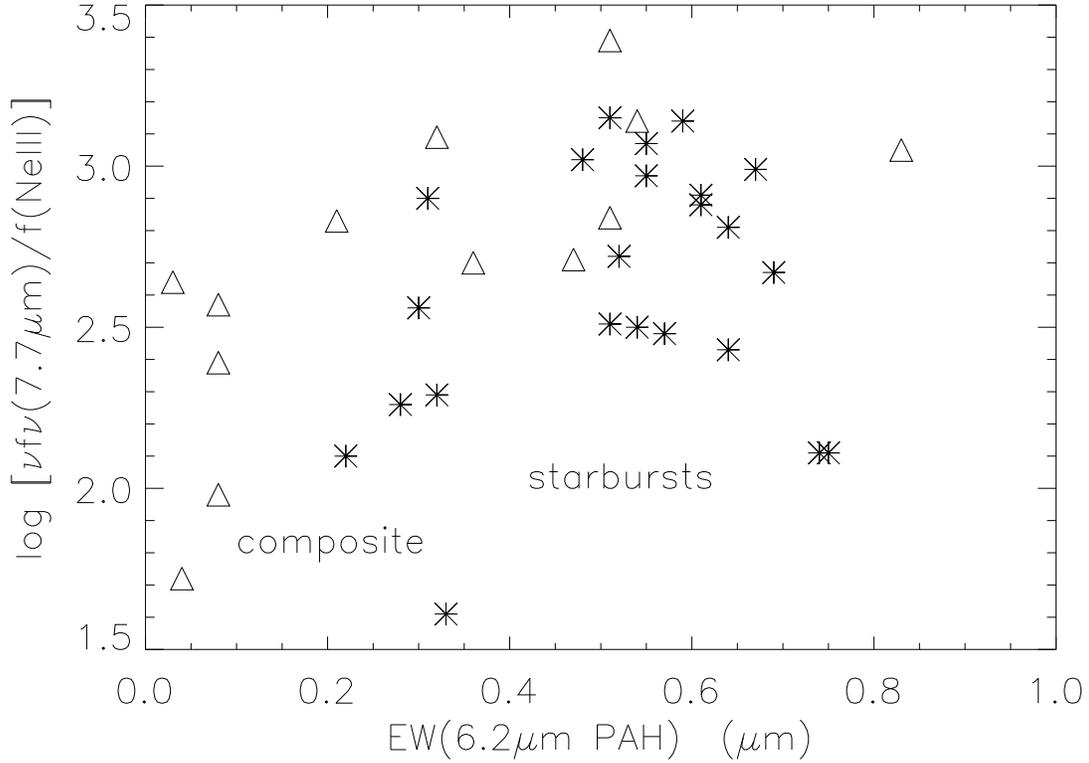}
\caption{Strength of PAH 7.7$\mu$m feature measured as $\nu$f$_{\nu}$(7.7$\mu$m) compared to strength of [NeIII] 15.6$\mu$m emission line as function of PAH EW for PAH sources in Table 2.  Asterisks are sources with measured [NeIII] fluxes; diamonds are lower limits of ratio for sources without measurable [NeIII]. Distribution of PAH to Ne ratio illustrates dispersion in using PAH compared to Ne for measuring SFR. Vertical line is minimum EW for pure starbursts in Figure 6, so sources to right of line give measure of dispersion for starbursts without any AGN contribution; dispersion of 0.3 about the median value yields uncertainty in SFR by a factor of two for pure starbursts.} 
\end{figure}

\begin{figure}
\figurenum{13}
 
\includegraphics[scale=0.9]{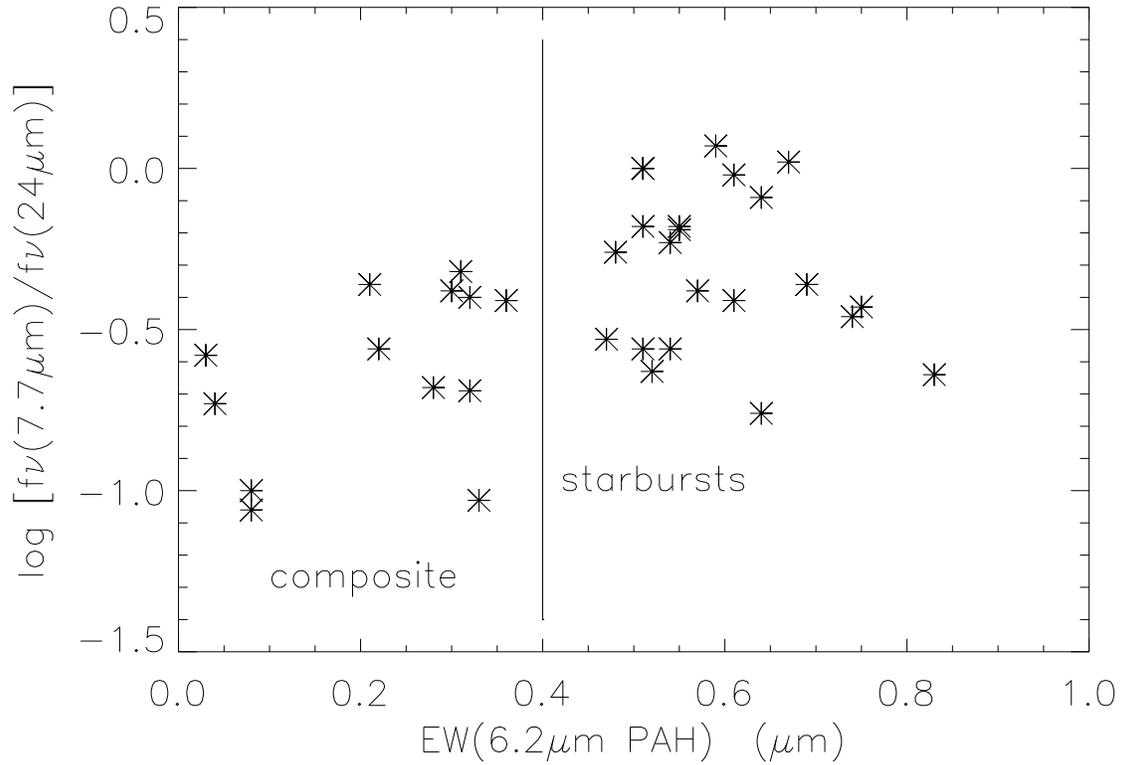}
\caption{Flux density of dust continuum at 24$\mu$m compared to flux density at peak of 7.7$\mu$m feature as function of PAH EW for PAH sources in Table 2. Distribution of dust to PAH ratio illustrates dispersion in using dust continuum compared to PAH for measuring SFR.  Vertical line is minimum EW for pure starbursts in Figure 6, so sources to right of line give measure of dispersion for starbursts without any AGN contribution; dispersion of 0.3 about the median value yields uncertainty in SFR by a factor of two for pure starbursts.} 
\end{figure}

\begin{figure}
\figurenum{14}
 
\includegraphics[scale=0.9]{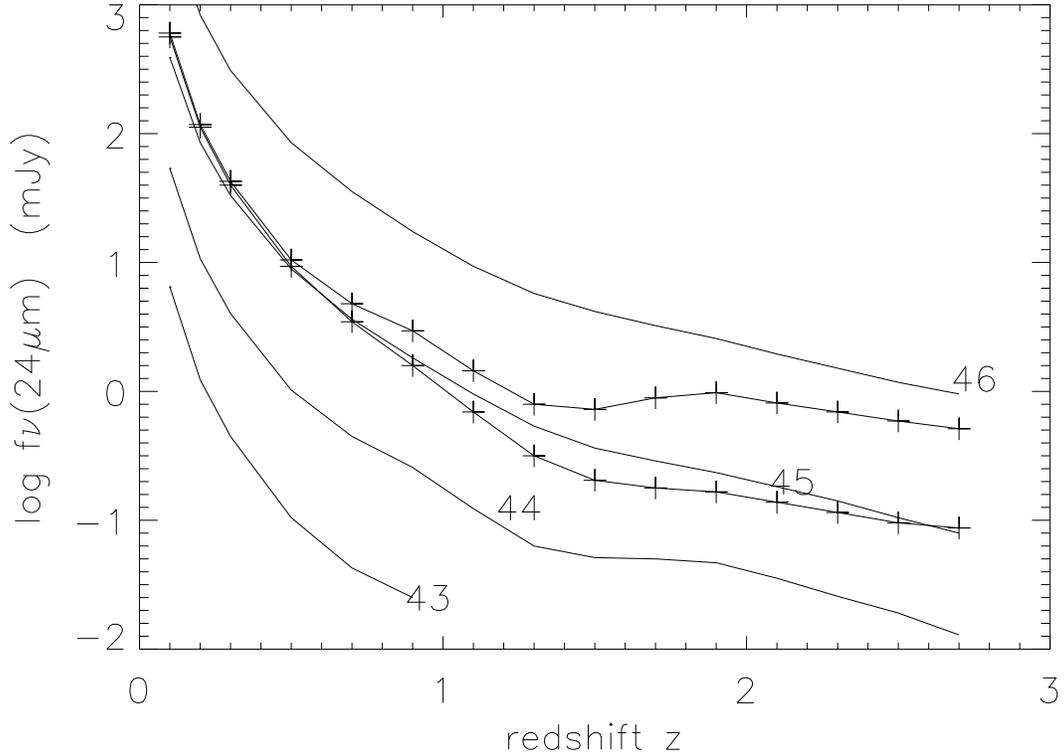}
\caption{Predicted flux densities f$_{\nu}$(24$\mu$m) which would be observed by MIPS as a function of redshift using average spectra of sources with different luminosities (solid lines). From top to bottom, luminosities and spectra used for curves are: log $\nu$L$_{\nu}$(15$\mu$m) = 46.0 using average spectrum for log $\nu$L$_{\nu}$(15$\mu$m) $>$ 45.0;  log $\nu$L$_{\nu}$(15$\mu$m) = 45.0 using average spectrum for 44.0 $<$ log $\nu$L$_{\nu}$(15$\mu$m) $<$ 45.0; log $\nu$L$_{\nu}$(15$\mu$m) = 44.0 using average spectrum for 43.0 $<$ log $\nu$L$_{\nu}$(15$\mu$m) $<$ 44.0; and log $\nu$L$_{\nu}$(15$\mu$m) = 43.0 using average spectrum for 42.0 $<$ log $\nu$L$_{\nu}$(15$\mu$m)$<$ 43.0. Curve for log $\nu$L$_{\nu}$(15$\mu$m) = 44.0 increases relative to brighter curves for z $>$ 1.5 because of strong PAH features in the average spectrum. Connected crosses show envelope that would arise for log $\nu$L$_{\nu}$(15$\mu$m) = 45.0 using shapes of extreme spectra in Figure 10, normalizing each spectrum to log $\nu$L$_{\nu}$(15$\mu$m) = 45.0.  Upper envelope illustrates that a source of this luminosity with strong PAH features would have f$_{\nu}$(24$\mu$m) $\sim$ 1 mJy at z $\sim$ 2, typical of numerous such sources that have been discovered in $Spitzer$ surveys. }  
\end{figure}

\end{document}